\documentclass{aa}  

\usepackage{graphicx}
\usepackage{txfonts}
\usepackage{multirow}
\usepackage{colortbl}

\usepackage{hyperref}  
\hypersetup{colorlinks=true,linkcolor=[rgb]{1.,0.2,0.2},citecolor=[rgb]{0.1,0.4,1.},filecolor=[rgb]{0.7,0.2,0.2},urlcolor=[rgb]{0.7,0.2,0.2}}

\usepackage{color}
\definecolor{blue}{rgb}{0., 0., 1}
\definecolor{lightblue}{rgb}{0.1,0.4,1.}

\newcommand{\ci}[1]{{\color{lightblue}{#1}}}
\newcommand {\CL}{A2744}
\newcommand {\LT}{\texttt{LensTool}}
\newcommand {\SLOT}{\texttt{SLOT}}
\newcommand {\T}{Table\,}

\newcommand {\Fig}{Fig.\,}
\newcommand {\Eq}{Eq.\,}

\newcommand {\ppxf}{\texttt{Ppxf}}

\defcitealias{Bergamini_2021}{B21}
\defcitealias{Richard_2021}{R21}

\begin{document} 
\title{New high-precision strong lensing modeling of Abell 2744}
\subtitle{Preparing for JWST observations}

\author{
P.~Bergamini \inst{\ref{unimi}, \ref{inafbo}} \fnmsep\thanks{E-mail: \href{mailto:pietro.bergamini@unimi.it}{pietro.bergamini@unimi.it}}
\and
A.~Acebron \inst{\ref{unimi},\ref{inafmilano}} \and
C.~Grillo \inst{\ref{unimi},\ref{inafmilano}} \and
P.~Rosati \inst{\ref{unife},\ref{inafbo}} \and
G.~B.~Caminha \inst{\ref{max_plank}} \and
A.~Mercurio \inst{\ref{inafna}} \and
E.~Vanzella \inst{\ref{inafbo}} \and
G.~Angora \inst{\ref{unife},\ref{inafna}} \and
G.~Brammer \inst{\ref{dawn},\ref{niels}} \and
M.~Meneghetti \inst{\ref{inafbo}} \and
M.~Nonino \inst{\ref{inafts}}
}
\institute{
Dipartimento di Fisica, Universit\`a  degli Studi di Milano, via Celoria 16, I-20133 Milano, Italy \label{unimi}
\and
INAF -- OAS, Osservatorio di Astrofisica e Scienza dello Spazio di Bologna, via Gobetti 93/3, I-40129 Bologna, Italy \label{inafbo} 
\and
INAF - IASF Milano, via A. Corti 12, I-20133 Milano, Italy
\label{inafmilano}
\and
Dipartimento di Fisica e Scienze della Terra, Universit\`a degli Studi di Ferrara, via Saragat 1, I-44122 Ferrara, Italy \label{unife}
\and
Max-Planck-Institut f\"ur Astrophysik, Karl-Schwarzschild-Str. 1, D-85748 Garching, Germany \label{max_plank}
\and
INAF -- Osservatorio Astronomico di Capodimonte, Via Moiariello 16, I-80131 Napoli, Italy \label{inafna}
\and
Cosmic Dawn Center (DAWN), Jagtvej 128, DK2200 Copenhagen N, Denmark \label{dawn}
\and
Niels Bohr Institute, University of Copenhagen, Jagtvej 128, København N, DK-2200, Denmark  \label{niels}
\and
INAF -- Osservatorio Astronomico di Trieste, via G. B. Tiepolo 11, I-34143, Trieste, Italy \label{inafts}
\and
INAF -- Osservatorio Astrofisico di Arcetri, Largo E. Fermi, I-50125, Firenze, Italy \label{arcetri}
           }

   \date{Received February 14, 2020; accepted February 14, 2020}

  \abstract
  {
We present a new strong lensing model of the Hubble Frontier Fields galaxy cluster Abell 2744, at $z=0.3072$, by exploiting archival Hubble Space Telescope (HST) multi-band imaging and Multi Unit Spectroscopic Explorer (MUSE) follow-up spectroscopy. The lens model considers 90 spectroscopically confirmed multiple images (from 30 background sources), which represents the largest secure sample for this cluster field prior to the recently acquired James Webb Space Telescope observations. The inclusion of the sub-structures within several extended sources as model constraints allows us to accurately characterize the inner total mass distribution of the cluster and the position of the cluster critical lines. We include the lensing contribution of 225 cluster members, 202 of which are spectroscopically confirmed. We complement this sample with 23 photometric member galaxies which are identified with a convolution neural network methodology with a high degree of purity.
We also measure the internal velocity dispersion of 85 cluster galaxies, down to $m_{F160W} = 22$, to independently estimate the role of the subhalo mass component in the lens model.
We investigate the effect of the cluster environment on the total mass reconstruction of the cluster core with two different mass parameterizations. We consider the mass contribution from three external clumps, either based on previous weak-lensing studies, or extended HST imaging of luminous members around the cluster core. 
In the latter case, the observed positions of the multiple images are better reproduced, with a remarkable accuracy of $\sim0.37\arcsec$, a factor of $\sim2$ smaller than previous lens models, that exploited the same HST and MUSE data-sets.
As part of this work, we develop and make publicly available a Strong Lensing Online Tool (\texttt{SLOT}) to exploit the predictive power and the full statistical information of this and future models, through a simple graphical interface. We plan to apply our new high-precision strong lensing model to the first analysis of the GLASS-JWST-ERS program, specifically to measure the intrinsic physical properties of high-$z$ galaxies from robust magnification maps.
}


   \maketitle


\section{Introduction}

\begin{figure*}
	\centering
	\includegraphics[width=1\linewidth]{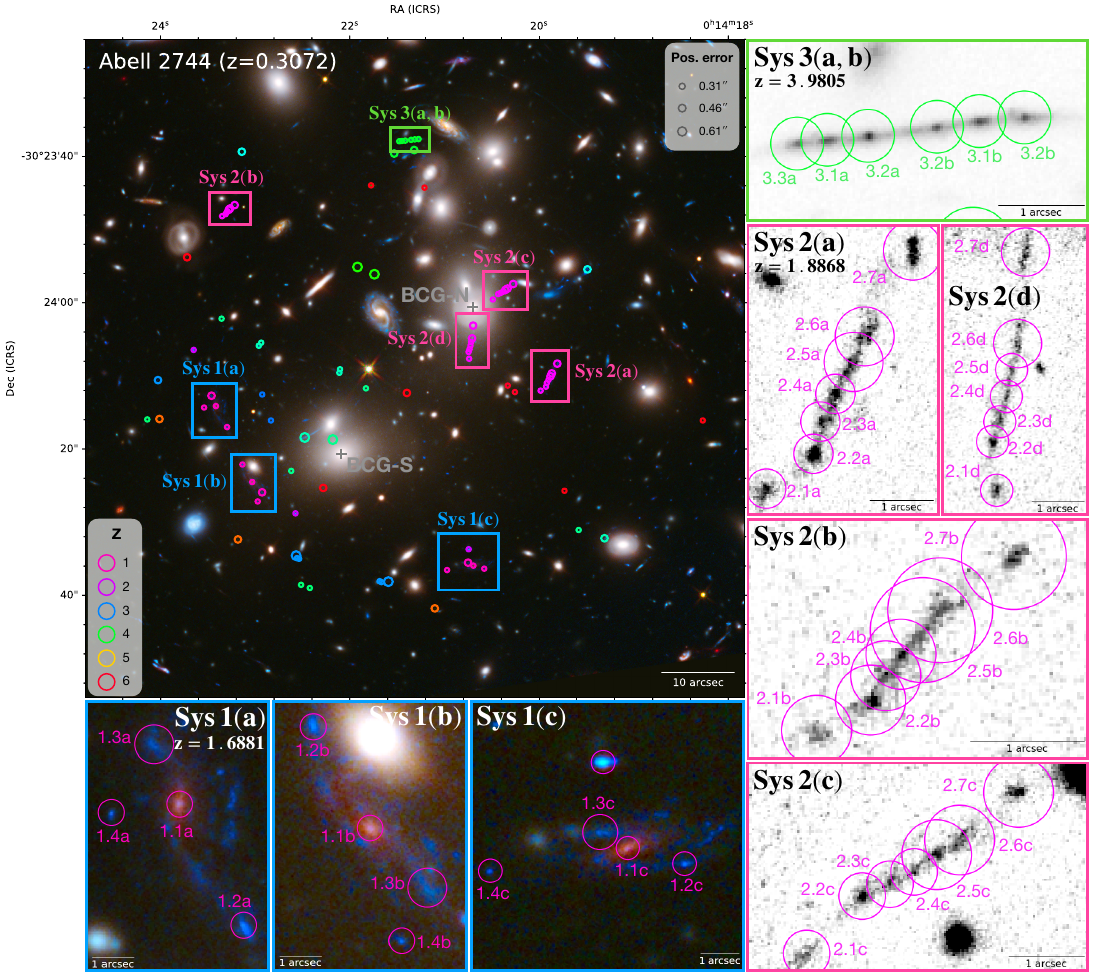}
	\caption{Color-composite RGB image of \CL\ (credits: \href{https://esahubble.org/images/heic1401a/}{NASA/ESA}). Circles show the positions of the 90 spectroscopically confirmed multiple images included in the SL model, color-coded according to their redshift value. The size of the circle illustrates the adopted (re-scaled) positional error in the modeling. Colored squares highlight the systems of multiple images for which additional lensed clumps have been identified. The two cluster BCGs (BCG-N and BCG-S) are labeled in gray.}
	\label{fig:RGB}
\end{figure*}

\begin{figure}
	\centering
	\includegraphics[width=1\linewidth]{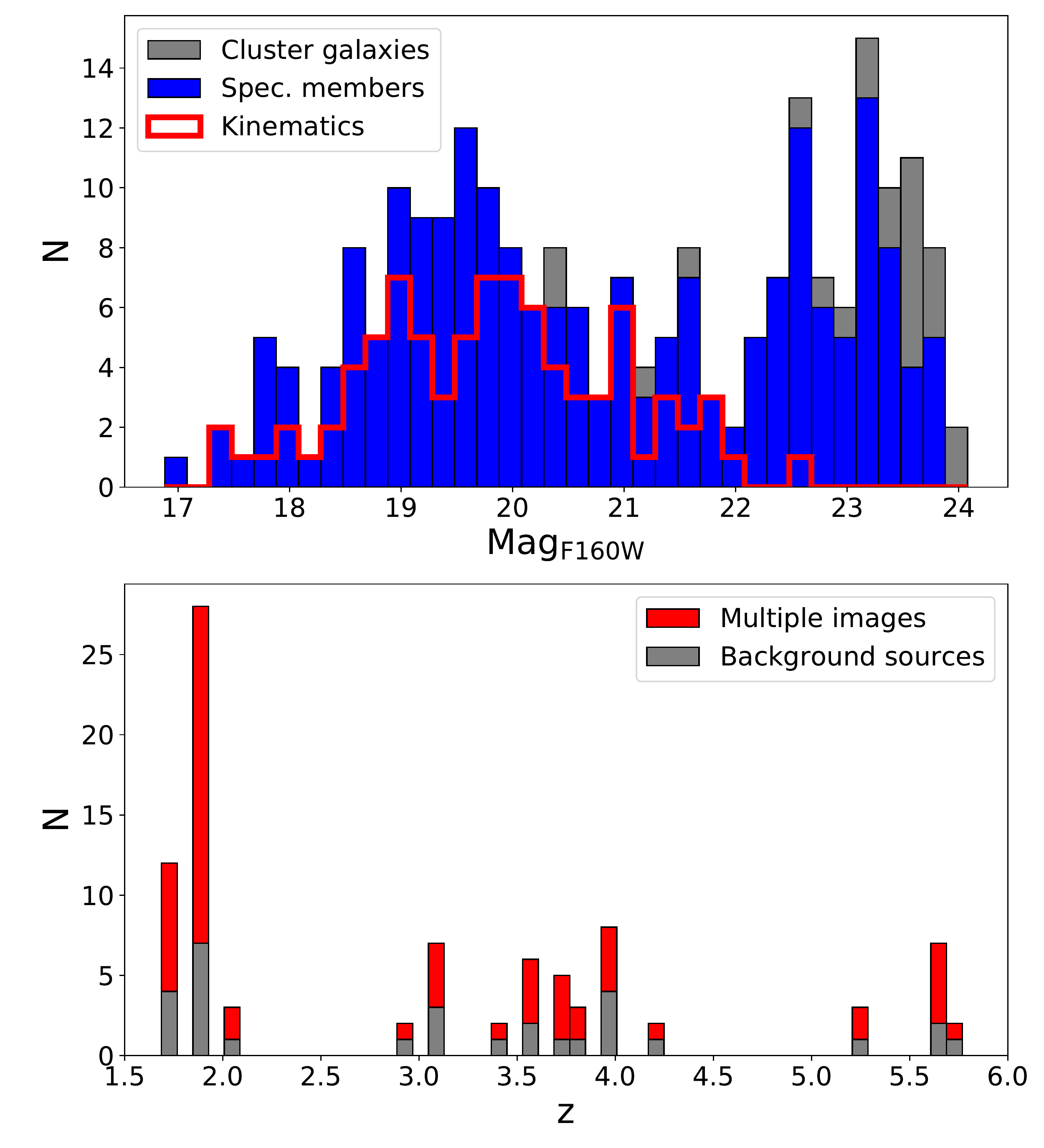}
	\caption{\textit{Top:} Distribution of cluster member galaxies as a function of their magnitudes in the HST/F160W filter. The photometric sample of cluster members used in our lens model is plotted in gray, with the spectroscopic members pictured in blue. Cluster members with a reliable measurement of their internal stellar velocity dispersion are highlighted in red.
	\textit{Bottom:} Redshift distribution of the observed 90 multiple images used to constrain the reference lens model described in this work.}
	\label{fig:hist}
\end{figure}

\begin{figure}
	\centering
	\includegraphics[width=1\linewidth]{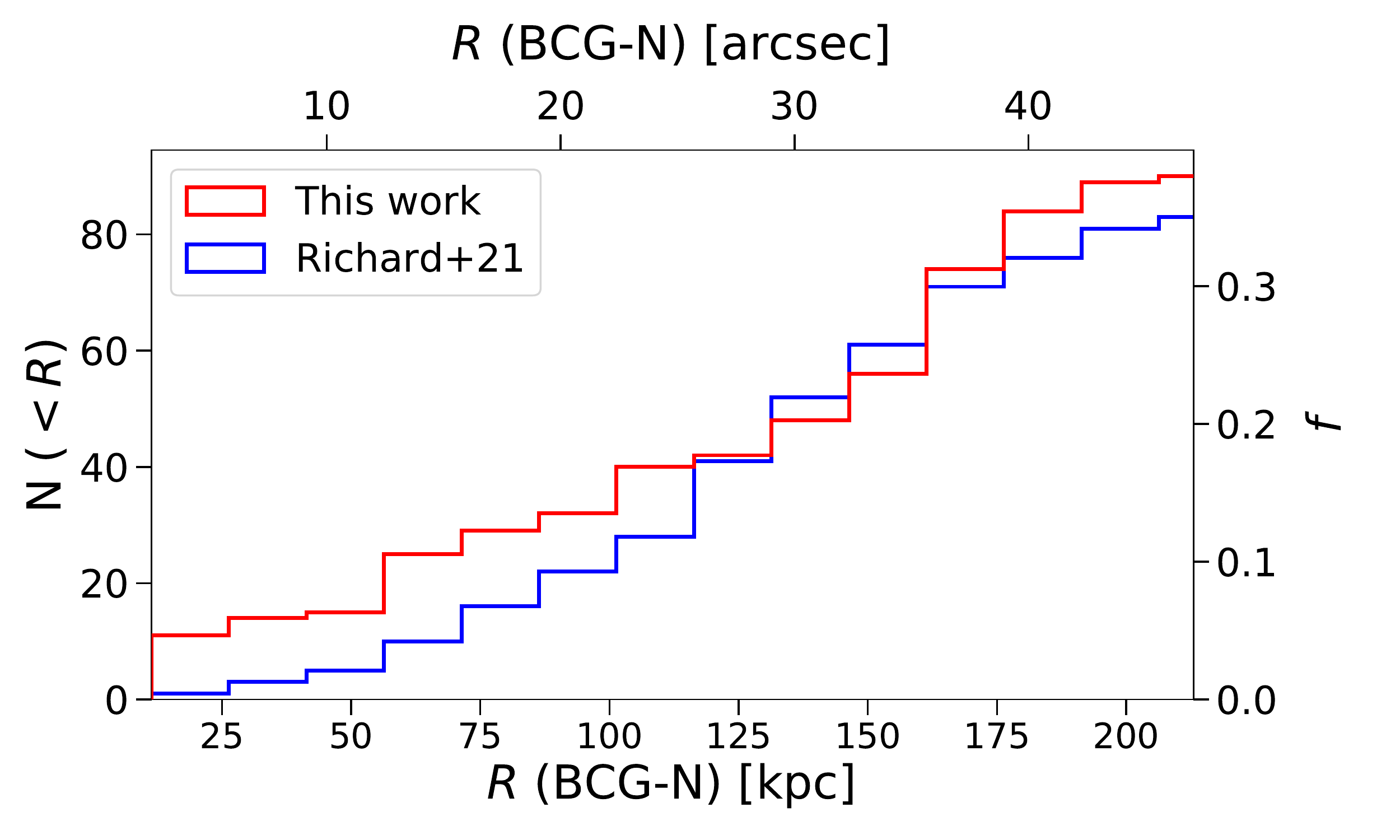}
	\caption{Cumulative distributions of the distances of the multiple images from the Northern BCG (BCG-N) of \CL: in red the distribution of the images used as constraints in this work (90 multiple images in total) and in blue the images used in \citetalias{Richard_2021}.}
	\label{fig:image_distribution}
\end{figure}

\begin{figure}
	\centering
	\includegraphics[width=1\linewidth]{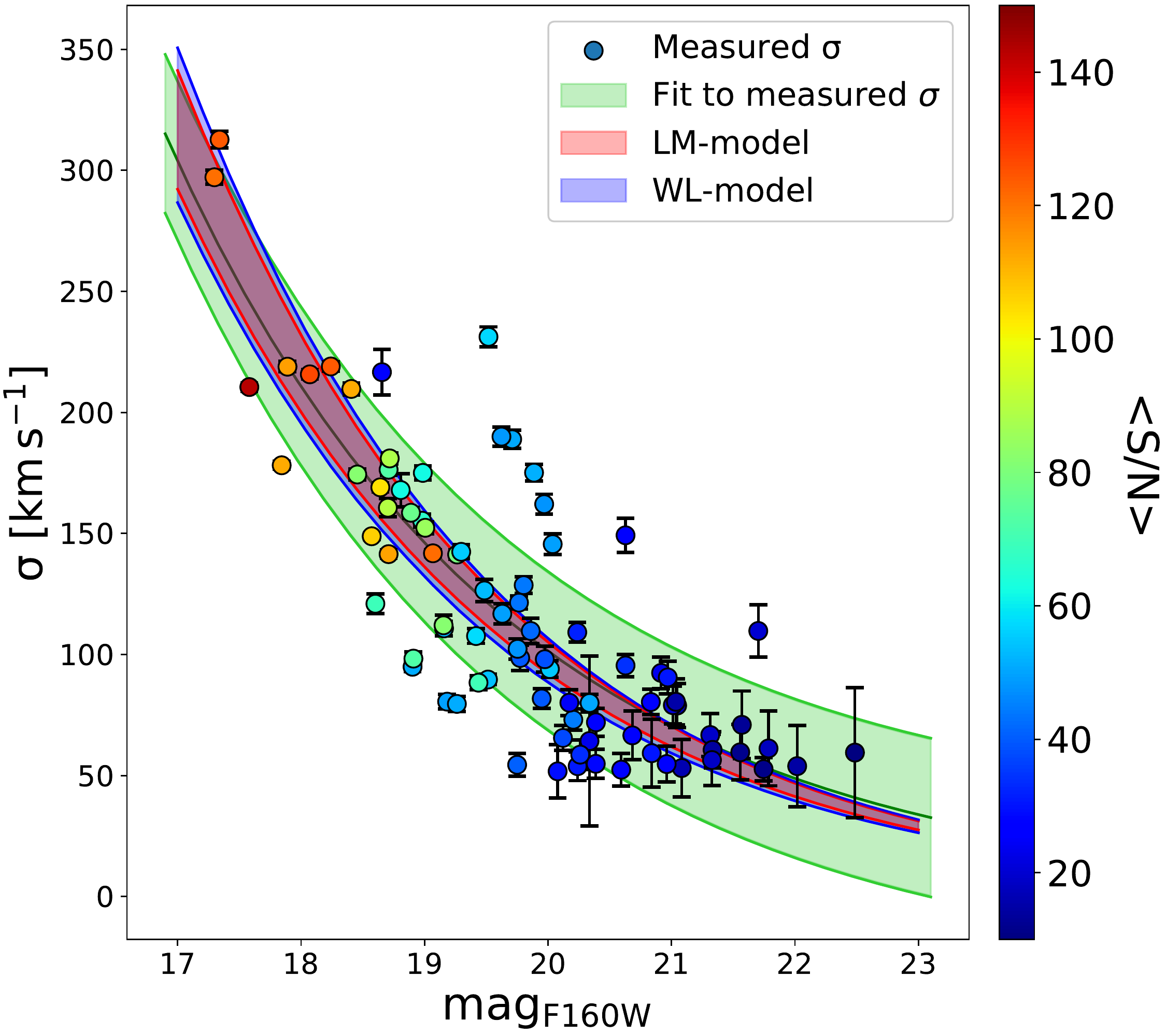}
	\caption{Measured internal stellar velocity dispersions of 85 cluster galaxies as a function of their magnitudes in the HST/F160W band are shown as filled circles, color-coded according to the mean signal-to-noise ratio of the galaxy spectra. The green line and filled area correspond to best-fit and the associated mean scatter of the $\sigma - m_{F160W}$ relation, respectively (see Section \ref{sec:cluster_members}). The red and blue areas correspond to the $68\%$ confidence level of the $\sigma - m_{F160W}$ relation obtained from the reference \texttt{LM-model} and \texttt{WL-model}, respectively.}
	\label{fig:green_plot}
\end{figure}

\begin{figure}
	\centering
	\includegraphics[width=1\linewidth]{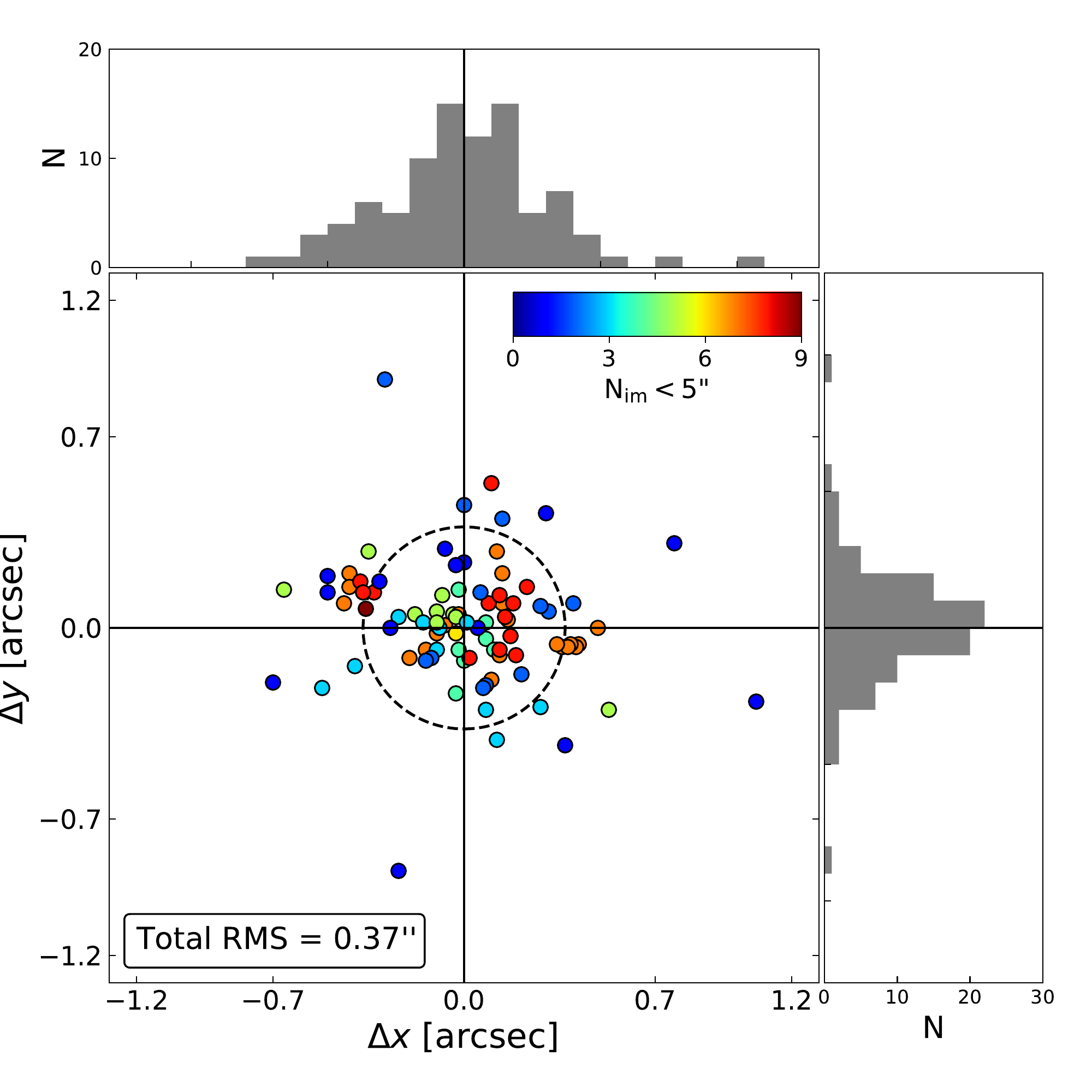}
	\caption{Displacements $\boldsymbol{\Delta}_i$ (see Eq. \ref{eq.: rms_lt}) along the $x$ and $y$ directions of the 90 observed multiple images used to optimize the reference lens model described in this work, color-coded according to the spatial density of the images within $5\arcsec$. The dashed black circle indicates the total $\Delta_{rms}$ value of $0.37\arcsec$. The histograms show the displacement distribution along each direction, also illustrating the goodness of the model.}
	\label{fig:rms}
\end{figure}

\begin{figure*}
	\centering
	\includegraphics[width=1\linewidth]{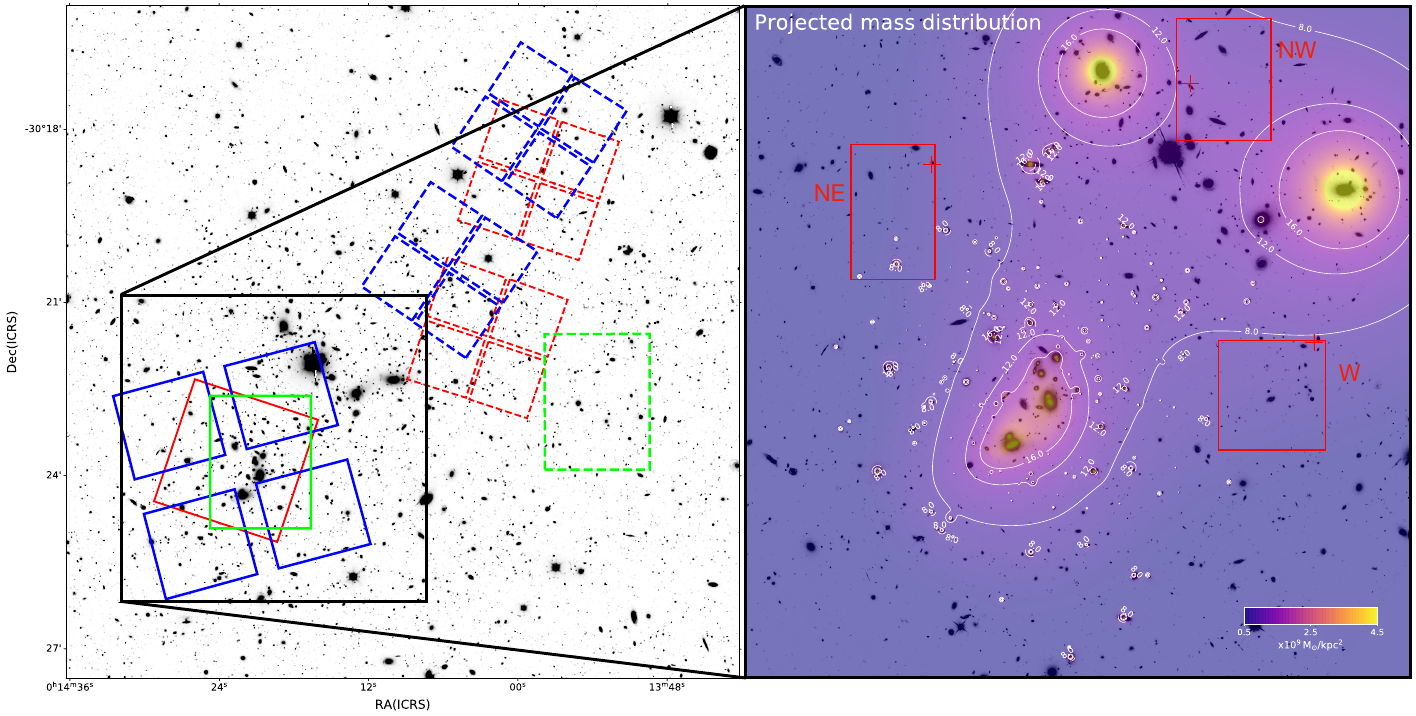}
	\caption{\textit{Left:} Magellan R-band imaging of \CL\ with the JWST footprints from the GLASS-JWST-ERS program, as also shown in Figure 1 from \citet{Treu2022}. In red and blue we show the footprints of the NIRISS and NIRSpec pointings, respectively, while in green we plot the HFF central pointing of the cluster and parallel field (dashed). Dashed lines correspond to parallel NIRCam pointings. \textit{Right:} Total projected mass distribution obtained from the best-fit \texttt{LM-model}, in units of $\rm 10^9~M_{\odot}~kpc^{-2}$, overlaid on the HST/F814W image. The red rectangles indicate the assumed priors for the positions of the three WL clumps from \citet{Medezinski2016}, while the red crosses show the obtained best-fit positions.}
	\label{fig:Mass_image}
\end{figure*}

\begin{figure}
	\centering
	\includegraphics[width=1\linewidth]{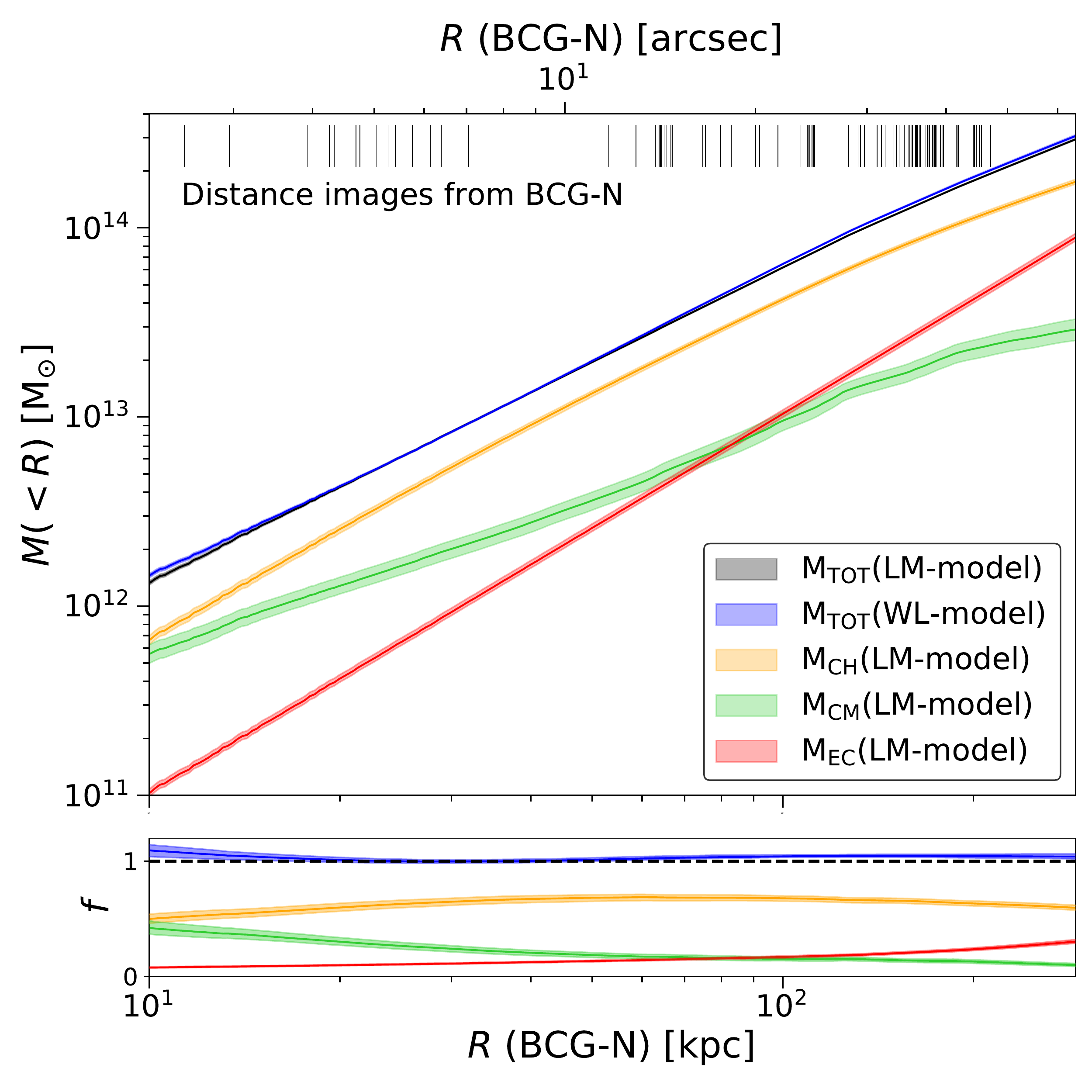}
	\caption{{\it Top:} Cumulative total mass distribution of \CL\ as a function of the projected distance $R$ from the northern BCG obtained from the reference lens model. The different components of the total mass are shown: cluster halo (CH), cluster members (CM, including the BCG), and external clump (EC).  The distances of the observed multiple images from the BCG-N are plotted using small vertical bars. {\it Bottom:} Ratio between the projected mass profiles obtained for the different color-coded mass components and the total cluster mass.}
	\label{fig:mass_profile}
\end{figure}

\begin{figure*}
	\centering
	\includegraphics[width=1\linewidth]{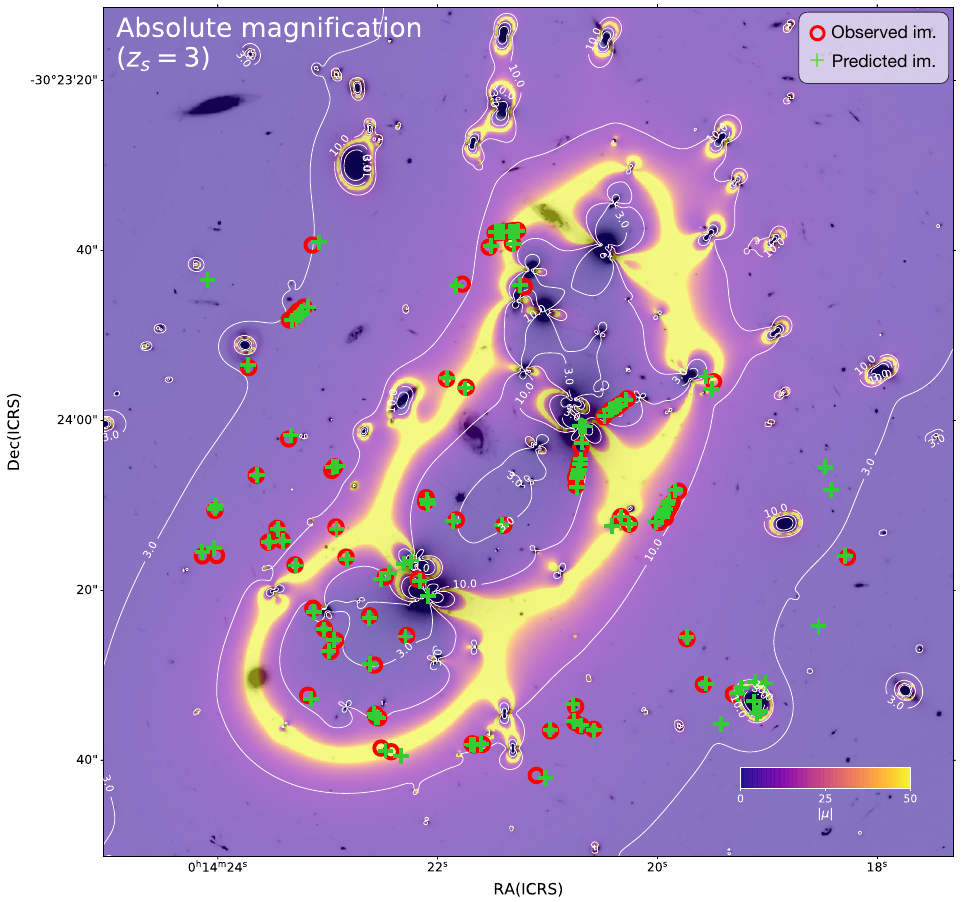}
	\caption{Absolute magnification map computed at $z_{\rm s}=3$. The red circles indicated the observed positions of the 90 multiple images, while the green crosses show the predicted positions obtained with our best-fit \texttt{LM-model}.}
	\label{fig:Magnification_image}
\end{figure*}

The combination of \textit{Hubble Space Telescope} (HST) high-resolution imaging with mainly ground-based spectroscopic follow-up of lens galaxy clusters has enabled a broad range of science cases, from the charaterization of the dark matter distribution in cluster cores \citep{Grillo_2015, Limousin_2016, Cerny_2018, Diego2020} to cluster physics \citep{Bonamigo_2017, Bonamigo_2018, Annunziatella_2017, Montes2022}, from cluster galaxy evolution \citep{Annunziatella2016, Mercurio2021} and the study of high-redshift galaxies \citep{Coe_2013, Vanzella_2020, Mestric2022} to cosmological analyses \citep{Jullo2010, Caminha_rxc2248, Grillo_2018}.
This has motivated numerous imaging programs with HST, such as the Cluster Lensing and Supernova survey with Hubble \citep[CLASH,][]{Postman_2012_clash}, the Hubble Frontier Fields program \citep[HFF,][]{Lotz_2017HFF}, the REionization LensIng Cluster Survey \citep[RELICS,][]{Coe_2019} and the Beyond Ultra-deep Frontier Fields And Legacy Observations \citep[BUFFALO,][]{Steinhardt2020} surveys. 
In parallel, follow-up spectroscopic campaigns have allowed to build high-precision and accurate strong lensing mass models \citep[e.g.,][]{Grillo2016, Caminha_2019, Lagattuta_2019, Bergamini_2021}, impacting directly the precision and accuracy of subsequent cluster lensing applications \citep[][]{Meneghetti_2020, Grillo_2018, Vanzella_2020,2022arXiv220409065M}.  
The advent of the James Webb Space Telescope (JWST) will push to new frontiers the studies mentioned above.
In this context, the GLASS James Webb Space Telescope Early Release Science program \citep[hereafter GLASS-JWST-ERS; JWST-ERS-1324: PI Treu,][]{Treu2022} has recently obtained the deepest ERS data, by pointing at the galaxy cluster Abell 2744.
 
Abell 2744 (\CL\ hereafter, see Figure \ref{fig:RGB}), at a redshift of $z=0.3072$, is a massive, X-ray luminous, merging galaxy cluster \citep{Allen1998, Ebeling2010} that has been the target of extensive multi-wavelength observations.
The detection of a central radio halo, and a large radio-relic in the North-East region of the cluster, led to the classification of \CL\ as a recent merging system \citep{Giovannini1999, Govoni2001}. Subsequent XMM-\textit{Newton} and \textit{Chandra} X-ray observations, combined with rich optical spectroscopy, revealed numerous substructures in the cluster field \citep{Kempner2004, Braglia2009, Owers2011, Eckert2015}. In addition, studies on the spatial distribution and kinematics of member galaxies suggest a complex internal structure in \CL\ \citep[see for instance][]{Couch1987, Girardi2001, Braglia2009}, showing a bimodal velocity distribution of member galaxies, with a high velocity component \citep[][]{Owers2011}.
Following the detection of the first strong lensing (SL) features in the core of the cluster \citep{Smail1997}, \CL\ has also been the subject of numerous lensing analyses, from SL free-form \citep[][]{Lam2014, Wang2015} and parametric \citep[][hereafter \citetalias{Richard_2021}]{Johnson_2014, Jauzac_2015_A2744, Mahler_2018, Richard_2021}, weak-lensing \citep[WL,][]{Medezinski2016}, to joint SL+WL models \citep[][]{Merten2011, Jauzac2016}. 
Due to its lensing strength, \CL\ was included as one of the six galaxy clusters in the HFF program with HST \citep{Lotz_2017HFF}, collecting some of the deepest high-resolution imaging of a cluster field.
While the HFF observations led to the identification of a very large number of photometric multiple images \citep[up to $\sim180$, see][]{Jauzac_2015_A2744}, the sample of secure systems remained fairly small, which has been shown to potentially introduce biases in total mass reconstructions \citep[][]{Grillo_2015, johnson17}. 
In particular, thanks to spectroscopic follow-up observations within the Grism Lens-Amplified Survey from Space survey \citep[GLASS,][]{Treu_2015, Schmidt_2014},  \citet{Wang2015} provided spectroscopic redshift measurements for 8 background sources. 
The avenue of MUSE \citep[Multi Unit Spectroscopic Explorer,][]{Bacon_MUSE} follow-up observations of \CL, combined with deep HFF imaging, enabled the number of spectroscopically confirmed multiple images to significantly rise \citep{Mahler_2018, Richard_2021}, leading to more accurate cluster mass models.

In this work, we further exploit archival high-resolution HST imaging and MUSE spectroscopy to build an improved strong lensing model of \CL. The new model includes the largest set of spectroscopically confirmed multiple images obtained so far in this cluster field and internal kinematics of cluster galaxies to independently constrain the subhalo total mass component. 
The new sample of multiple images consists of multiply lensed clumps within resolved extended sources. These additional systems are especially efficient at constraining tightly the position of the critical lines locally \citep[see for instance][]{Grillo2016, Bergamini_2021}.

The paper is organized as follows: Section \ref{sec:Data} describes the HST imaging and spectroscopic data-sets used to develop the new lens model of \CL. In Section \ref{sec:model_description}, we detail the adopted methodology for the strong lens modeling, together with the selection of the multiple images and cluster members. Our results and strong lensing online tool are presented in Sections \ref{sec:results} and \ref{sec:SLOT}, respectively, and our main conclusions are drawn in Section \ref{sec:conclusions}.

Throughout this work, we adopt a flat $\Lambda$CDM cosmology
with $\Omega_m = 0.3$ and $H_0= 70\,\mathrm{km\,s^{-1}\,Mpc^{-1}}$. Using this cosmology, a projected distance of $1\arcsec$ corresponds to a physical scale of 4.528 kpc at the \CL\ redshift of $z=0.3072$. All magnitudes are given in the AB system.


\section{Data}
\label{sec:Data}
This section presents the photometric and spectroscopic data sets used in this work.

\subsection{HST imaging}
\label{sec:dataHST}
As part of the Hubble Frontier Fields program \citep[HFF, Proposal ID: 13495,][]{Lotz_2017HFF}, \CL\ is one of the cluster fields with the deepest high-resolution observations obtained with HST. The cluster was imaged, from October 2013 to July 2014, in the optical and near-infrared with seven different bands from the Advanced Camera for Surveys (ACS; F435W, F606W, F814W) and the Wide Field Camera 3 (WFC3; F105W, F110W, F140W, F160W). 
The Frontier Fields observations, including ancillary data from previous HST programs with the same filters, were reduced with the HST science data products pipeline \citep{Koekemoer_2014HFF}.
The HFF obervations of \CL\ were recently extended out to a larger radius, thanks to the BUFFALO survey \citep[][]{Steinhardt2020} that has provided shallower imaging of the outskirts of the six HFF clusters.
In the following analysis, we focus on the core of the galaxy cluster and defer for future work an extended strong lensing analysis.
We thus exploit the Frontier Fields HST mosaics with a pixel scale value of $0\arcsec.03$\footnote{\url{https://archive.stsci.edu/prepds/frontier/}}. 

\subsection{VLT/MUSE \& ancillary spectroscopy}
\label{sec:dataMUSE}
\CL\ has also been the target of extensive spectroscopic campaigns with several instruments.
In particular, we use archival observations from the MUSE integral field spectrograph, mounted on the VLT \citep[Very Large Telescope, ][]{Bacon_MUSE}, obtained within the GTO Program 094.A-0115 (PI: Richard). The data, consisting of four MUSE pointings, are described in \citet{Mahler_2018} (see Figure 1 for the exposure time within each MUSE pointing), while the reduction process is presented in \citetalias{Richard_2021}.
The MUSE data cube is reduced and analyzed following the procedure adopted in \citet{Caminha_macs0416, Caminha_macs1206, Caminha_2019}, using the standard reduction pipeline \citep[version 2.8.5,][]{Weilbacher2020}. In addition, we use the \emph{auto-calibration} method and the Zurich Atmosphere Purge \citep[ZAP,][]{Soto2016} to improve the data reduction. The data have a FWHM value of $0\arcsec.61$.
We proceed to remeasure the redshifts of objects classified as either cluster members or multiple images in \citetalias{Richard_2021}. The 1-dimensional spectra of these objects are extracted within a $0.\arcsec8$ radius circular aperture, while we apply custom apertures for faint sources, based on their estimated morphology from the HST imaging. We exploit spectral templates, as well as the identification of emission lines to build our redshift catalogs. The reliability of each redshift measurement is then quantified with the following Quality Flag (QF) assignments: \textit{insecure} (QF = 1), \textit{likely} (QF = 2), \textit{secure} (QF = 3), and \textit{based on a single emission line} (QF = 9).

In addition, \CL\ was targeted for 4.4h with the wide-field multi-object spectrograph VIMOS (VIsible Multi-Object Spectrograph) betweeen August 14-16, 2004 as part of the ESO Large Program 169.A-0595 (PI: Böhringer). The spectroscopic catalog, presented in \citet{Braglia2009}, includes 395 non-stellar objects with a spectroscopic confirmation.
The cluster was again observed using the AAOmega multi-object spectrograph on the 3.9m Anglo-Australian Telescope (AAT) between 12–18 September 2006. Combined with previous (in particular the VIMOS catalog), \citet{Owers2011} provided spectroscopic redshifts measurements for 1237 non-stellar objects within $\sim15\arcmin$ of the cluster center, of which 343 were identified as cluster members.
Finally, the GLASS (Grism Lens Amplified Survey from Space) HST WFC3/IR grism GO program\footnote{\url{archive.stsci.edu/prepds/glass/}} \citep[][]{Treu_2015, Schmidt_2014} provided reliable redshift measurements for 81 non-stellar objects, with a quality flag of probable or secure.


\section{Strong lensing modeling}
\label{sec:model_description}
We develop a new lens model of \CL\ using the publicly available software \LT\footnote{\url{https://projets.lam.fr/projects/lenstool/wiki}}\ \citep{Kneib_lenstool, Jullo_lenstool, Jullo_Kneib_lenstool}, which reconstructs the total mass distribution of a galaxy cluster by exploiting a Bayesian technique. This code was very successful at reconstructing the mass distribution of several galaxy clusters and was among the best performing codes in the Frontier Fields Lens Modeling Comparison Project \citep{Meneghetti_2017}.
The best-fit parameters are found by minimizing a $\chi^2$ function, which quantifies the goodness of the lens model in reproducing the point-like positions of the observed multiple images. This statistics is defined as:

\begin{equation}
    \label{eq.: chi_lt}
    \chi^2(\pmb{\xi}) := \sum_{j=1}^{N_{fam}} \sum_{i=1}^{N_{im}^j} \left(\frac{\left\| \mathbf{x}_{i,j}^{obs} - \mathbf{x}_{i,j}^{pred} (\pmb{\xi}) \right\|}{\Delta x_{i,j}}\right)^2,
\end{equation}

\noindent where $\mathbf{x}_{i,j}^{obs}$ represents the observed position of the $i^{th}$ multiple image of the $j^{th}$ background source (images from the same source are called a family of multiple images), and $\mathbf{x}_{i,j}^{pred}$, its predicted position, given the set of model free parameters $\pmb{\xi}$. $\Delta x_{i,j}$ is the error associated to the position of the image.

While the best-fit model corresponds to the set of values of model free parameters for which the $\chi^2(\pmb{\xi})$ assumes its minimum value, we quote in the following the values for the parameters, and their associated errors, from the $50^{th}$, $16^{th}$, and $84^{th}$ percentiles of the marginalized posterior distributions.
Before sampling the posterior distributions, the initial positional uncertainty for each image, $\Delta x_{i,j}$, is re-scaled so that the $\chi^2$ value is close to the number of degrees of freedom in the model, defined as: $\mathrm{dof}=2\times [N^{tot}_{im} - N_{fam}] - N_{freepar}$.

Moreover, we also consider and quote the root-mean-square separation between the observed and model-predicted positions of the multiple images as another figure of merit to quantify the goodness of a lens model, which is estimated as:

\begin{equation}
    \label{eq.: rms_lt}
    \Delta_{rms}=\sqrt{\frac{1}{N_{im}^{tot}}\sum_{i=1}^{N_{im}^{tot}}\left\|\boldsymbol{\Delta}_i \right\|^2},
\end{equation}

\noindent where $\boldsymbol{\Delta}_i=\boldsymbol{x}_{i}^{obs} -\boldsymbol{x}_{i}^{pred}$ is the separation between the observed and predicted positions of the $i$-th image.

In this section, we present the catalog of multiple images used in the model optimization, the selection and stellar kinematic measurements of member galaxies, and a summary of the adopted mass parametrization \citep[see e.g., ][for a detailed description]{Bergamini_2019}.

\subsection{Multiple images}
\label{sec:multiple_images}
In this work, we consider previous identifications of multiple image systems presented in \citetalias{Richard_2021} and reanalyze the HST multi-band imaging and the MUSE data-cube (see Section \ref{sec:Data}). 
The selection of secure samples of multiple images is crucial when building accurate and high-precision cluster mass models, to avoid potential biases introduced by less reliable constraints \citep{Grillo_2015, johnson17}.
Therefore, we construct our sample by considering only secure systems, that are spectroscopically confirmed by our VLT/MUSE analysis with a QF value $\geq2$.
In addition, we introduce a \textit{Positional Quality Flag} (QP) that is then translated into different values for the initial positional uncertainty, $\Delta x_{i,j}$ in Equation \ref{eq.: chi_lt}, assumed in the lens model (see Figure \ref{fig:RGB}). Each image is given a value of QP=1 (compact HST emission), QP=2 (diffuse or elongated HST emission) or QP=3 (MUSE only detection).

As illustrated in Figure \ref{fig:hist} (bottom), the final sample of multiple image systems included in the lens model spans a large redshift range, between $z=1.69$ and $z=5.73$, with a total of 90 multiple images from 30 background sources (see Table \ref{table:lens_models}). This represents the largest spectroscopic sample of multiple images adopted so far for \CL. 
The multiple image positions are shown in Figure \ref{fig:RGB} and their properties are summarized in Table \ref{tab:multiple_images}. 
The observed image positions are used as constraints in the lens model, providing in total 180 observables and 60 free parameters for the positions of the corresponding sources.
All systems included in our lens model are discussed and compared below to the \textit{Gold} sample of \citetalias{Richard_2021}. The resulting cumulative distribution of the distances of the multiple images included in the lens models is shown in Figure \ref{fig:image_distribution}, and compared to that from \citetalias{Richard_2021}.

\textit{Systems 1, 2, 3, 4, and 26} appear as extended images, clearly showing several resolved emission regions. In in this work, we use as constraints all the multiply lensed clumps that can be securely identified. From these systems we build a total number of 118 observational constraints, compared to the 46 from \citetalias{Richard_2021}. This significant increase in the number of constraints in the innermost region of the cluster is illustrated in Figure \ref{fig:image_distribution}.
An example of the new identifications is highlighted in the zoom-in insets in Figure \ref{fig:RGB} for systems 1, 2 and 3. 

All images within \textit{Systems 6, 8, 18, 22, 34, 42, 62, 63 and 64} are included both in the catalog from \citetalias{Richard_2021} and ours, with no discrepancies in the redshift values.

\textit{Systems 5, 105, 47 and 147}, that form several extended images in the northernmost region of the cluster's core, are not included in our image sample. While we measure a redshift value in agreement with that from \citetalias{Richard_2021}, no clear counter image positions can be identified from the HST imaging.

\textit{Systems 10, 24, 30, 31, 41 and 61} have all a secure (QF=2 or QF=3) spectroscopic confirmation for only one of the multiple images. The remaining images have either a tentative (QF=1) or no redshift measurement. Therefore these systems cannot be considered in the secure sample. 

\textit{System 33} is composed of three multiple images. Images 33.1a and 33.1b are in common in the two catalogs (with no redshit discrepancy) while image 33.1c is not considered in ours as no redshift measurement is possible.  

\textit{System 37} has a redshift measurement for one of the two images included in \citetalias{Richard_2021}. This measurement was obtained with the Low Resolution Imager and Spectrograph (LRIS) at the Keck-I telescope \citep{Mahler_2018}. As the redshift cannot not be confirmed with MUSE, we do not include this system in our sample.

Finally, no redshift measurement was possible for \textit{Systems 39 and 40}. Thus, we remove them from our secure catalog.

\subsection{Cluster members selection \& stellar kinematics}
\label{sec:cluster_members}

Cluster member galaxies are selected based both on spectroscopic (see Section \ref{sec:dataMUSE}) and multi-band HST photometric (see Section \ref{sec:dataHST}) information.

Spectroscopically confirmed cluster members are identified as those galaxies, brighter than $m_{\rm F160W} = 24$, and with rest-frame ($z=0.3072$) relative velocities within $ 3000 ~ \rm km\,s^{-1}$, which corresponds to the redshift range [0.28-0.34]. 
We mainly exploit the MUSE data-cube to identify 162 galaxies with a reliable redshift estimate (i.e. with a QF $\geq$ 2). We also include member galaxies based on spectrocpic measurements from ancillary data sets with publicly available catalogs: 32 galaxies are securely identified from the AAT/AAOmega observations \citep{Owers2011}, 5 objects from GLASS \citep{Treu_2015, Schmidt_2014}, 2 from the VIMOS survey \citep{Braglia2009}, and 1 galaxy from \citet{Couch1987}.
28 spectroscopic galaxies from the ancillary AAT/AAOmega catalogue fall outside of the HST/WFC3 field of view. We therefore adopt the following color-magnitude relation to infer the F160W magnitudes:  $m_{F814}-m_{F160}=2.51-0.0797\times m_{F814}$.

We complete the spectroscopic sample by selecting 23 additional photometric, bright ($m_{\rm F160W} \leq 24$,) members based on a convolution neural network (CNN) technique, which identifies cluster members using multi-band HST image cutouts together with an extensive spectroscopic coverage, as part of the CLASH-VLT program combined with MUSE archival observations \citep[see][for a detailed description of the method]{Angora_2020}. The training set is composed of $\sim$3300 samples, in 14 CLASH and HFF clusters (with a redshift between $z=0.2 -0.6$). When tested on the spectroscopic sample of A2744, we measure a completeness level of 88\% and a high degree of purity, equal to 95\%.

In summary, our final high-purity cluster member catalog, which is integrated in the following lensing analysis, consists of 225 member galaxies in total, covering an area of $\sim 14 \rm ~ arcmin^{2}$. Within this sample, 202 (or $\sim 90\%$) are spectroscopically confirmed, and 23 are photometric members. We show in Figure \ref{fig:hist} the distribution of these cluster galaxies as a function of their magnitude in the HST/F160W band, and their properties are listed in Table \ref{tab:cluster_members}.

As presented in \citet{Bergamini_2019, Bergamini_2021}, we further exploit the MUSE data-cube to measure the line-of-sight stellar velocity dispersion for a large sub-set of cluster members. 
We extract the spectra for the 162 cluster galaxies securely confirmed by the MUSE data within $0.8''$ radius apertures (comparable to the MUSE PSF). Velocity dispersions are then measured using the publicly available software Penalized Pixel-Fitting method \citep[\ppxf,][]{Cappellari_2004, Cappellari_2017}, over the wavelength range [3700-5700] \r{A}.
In order to exploit reliable measurements in the subsequent lensing analysis, we limit the sample to galaxies with $\langle S/N \rangle >10$ and $\sigma_0 > 50~ \mathrm{km~s^{-1}}$ \citep[see e.g.,][]{Bergamini_2019, Bergamini_2021}.
In addition, we perform a visual inspection of the imaging and the fitted spectra, resulting in the exclusion of three faint galaxies whose spectra are contaminated by the light from BCG-S and BCG-N.
The final sample of cluster members with internal kinematics includes thus 85 galaxies, down to $m_{\rm F160W} \sim 22$ (see Figure \ref{fig:hist}, top). The resulting measured $\sigma_0$ values are presented in Figure \ref{fig:green_plot}, as a function of their HST/F160W magnitude values.
\subsection{Total mass parametrization}
\label{sec:MassParametrization}

\begin{table*}[]     
	\tiny
	\def\arraystretch{2.3}
	\centering          
	\begin{tabular}{|c|c|c|c|c|c|c|c|c|}
	    \cline{3-9}
		\multicolumn{2}{c|}{} & \multicolumn{7}{c|}{ \textbf{Input parameter values and assumed priors}} \\
		\cline{3-9}
		  \multicolumn{2}{c|}{} & \boldmath{$x\, \mathrm{[arcsec]}$} & \boldmath{$y\, \mathrm{[arcsec]}$} & \boldmath{$e$} & \boldmath{$\theta\ [^{\circ}]$} & \boldmath{$\sigma_{LT}\, \mathrm{[km\ s^{-1}]}$} & \boldmath{$r_{core}\, \mathrm{[arcsec]}$} & \boldmath{$r_{cut}\, \mathrm{[arcsec]}$} \\ 
          \hline

		  \multirow{5}{*}{\rotatebox[origin=c]{90}{\textbf{Cluster-scale halos}}} 
		  
		  & \boldmath{$1^{st}$} \bf{Cluster Halo} & $-5.0\,\div\,5.0$ & $-5.0\,\div\,5.0$ & $0.0\,\div\,0.9$ & $0.0\,\div\,180.0$ & $300.0\,\div\,1500.0$ & $0.0\,\div\,30.0$ & $2000.0$ \\
		  
		  & \boldmath{$2^{nd}$} \bf{Cluster Halo} & $-27.9\,\div\,-7.9$ & $-30.1\,\div\,-10.1$ & $0.0\,\div\,0.9$ & $0.0\,\div\,90.0$ & $300.0\,\div\,1500.0$ & $0.0\,\div\,30.0$ & $2000.0$  \\
		  
		  \cline{2-9}

		  & \boldmath{$1^{st}$} \bf{Ext. clump} & $99.5$ & $86.0$ & $0.0$ & $0.0$ & $100.0\,\div\,1500.0$ & $0.001$ & $2000.0$  \\
		  
		  & \boldmath{$2^{nd}$} \bf{Ext. clump} & $138.3$ & $99.9$ & $0.0$ & $0.0$ & $100.0\,\div\,1500.0$ & $0.001$ & $2000.0$  \\
		  
		  & \boldmath{$3^{rd}$} \bf{Ext. clump} & $24.2$ & $155.8$ & $0.0$ & $0.0$ & $100.0\,\div\,1500.0$ & $0.001$ & $2000.0$  \\

          \hline
          \multicolumn{1}{c}{}
          \\[-5ex]

          \hline

		  \multirow{3}{*}{\rotatebox[origin=c]{90}{\textbf{Subhalos}}}

		  & \bf{BCG-N} & $0.0$ & $0.0$ & $0.0\,\div\,0.9$ & $0.0\,\div\,180.0$ & $200.0\,\div\,400.0$ & $0.0001$ & $0.1\,\div\,50.0$ \\
		  
  		  & \bf{BCG-S} & $-17.9$ & $-20.0$ & $0.0\,\div\,0.9$ & $0.0\,\div\,180.0$ & $200.0\,\div\,400.0$ & $0.0001$ & $0.1\,\div\,50.0$ \\
		  
		  \cline{2-9}
		  
		  & \bf{Scaling relations} & $\boldsymbol{N_{gal}=}223$
		  & $\boldsymbol{m_{\mathrm{F160W}}^{ref}=}17.34$
		  & $\boldsymbol{\alpha=}0.40$ & $\boldsymbol{\sigma_{LT}^{ref}=}190.0\,\div\,300.0$ & $\boldsymbol{\beta_{cut}=}0.41$ & $\boldsymbol{r_{cut}^{ref}=}0.5\,\div\,10.0$ & $\boldsymbol{\gamma=}0.20$\\
		  
		  \hline
		 
	\end{tabular}
	\\[6ex]
	\begin{tabular}{|c|c|c|c|c|c|c|c|c|}
	    \cline{3-9}
		\multicolumn{2}{c|}{} & \multicolumn{7}{c|}{ \textbf{Optimized output parameters}} \\
		\cline{3-9}
		  \multicolumn{2}{c|}{} & \boldmath{$x\, \mathrm{[arcsec]}$} & \boldmath{$y\, \mathrm{[arcsec]}$} & \boldmath{$e$} & \boldmath{$\theta\ [^{\circ}]$} & \boldmath{$\sigma_{LT}\, \mathrm{[km\ s^{-1}]}$} & \boldmath{$r_{core}\, \mathrm{[arcsec]}$} & \boldmath{$r_{cut}\, \mathrm{[arcsec]}$} \\ 
          \hline

		  \multirow{5}{*}{\rotatebox[origin=c]{90}{\textbf{Cluster-scale halos}}} 
		  
		  & \boldmath{$1^{st}$} \bf{Cluster Halo} & $-1.5^{+0.3}_{-0.4}$ & $-0.1^{+0.8}_{-0.9}$ & $0.6^{+0.1}_{-0.1}$ & $90.3^{+2.5}_{-2.7}$ & $522.7^{+32.1}_{-32.2}$ & $6.8^{+0.8}_{-0.7}$ & $2000.0$ \\
		  
		  & \boldmath{$2^{nd}$} \bf{Cluster Halo} & $-18.2^{+0.5}_{-0.5}$ & $-15.7^{+0.4}_{-0.4}$ & $0.4^{+0.1}_{-0.1}$ & $53.3^{+2.6}_{-2.8}$ & $633.9^{+21.1}_{-22.3}$ & $7.6^{+0.6}_{-0.6}$ & $2000.0$  \\
		  
		  \cline{2-9}

		  & \boldmath{$1^{st}$} \bf{Ext. clump} & $99.5$ & $86.0$ & $0.0$ & $0.0$ & $201.3^{+124.0}_{-73.8}$ & $0.001$ & $2000.0$  \\
		  
		  & \boldmath{$2^{nd}$} \bf{Ext. clump} & $138.3$ & $99.9$ & $0.0$ & $0.0$ & $933.4^{+32.3}_{-47.4}$ & $0.001$ & $2000.0$  \\
		  
		  & \boldmath{$3^{rd}$} \bf{Ext. clump} & $24.2$ & $155.8$ & $0.0$ & $0.0$ & $775.8^{+29.6}_{-30.8}$ & $0.001$ & $2000.0$  \\

          \hline
          \multicolumn{1}{c}{}
          \\[-5ex]

          \hline

		  \multirow{3}{*}{\rotatebox[origin=c]{90}{\textbf{Subhalos}}}

		  & \bf{BCG-N} & $0.0$ & $0.0$ & $0.3^{+0.2}_{-0.2}$ & $129.6^{+20.9}_{-25.4}$ & $221.8^{+13.0}_{-12.0}$ & $0.0001$ & $36.9^{+8.7}_{-10.5}$ \\
		  
  		  & \bf{BCG-S} & $-17.9$ & $-20.0$ & $0.8^{+0.1}_{-0.1}$ & $26.1^{+3.7}_{-3.0}$ & $304.4^{+9.4}_{-10.7}$ & $0.0001$ & $34.9^{+9.5}_{-9.6}$ \\
		  
		  \cline{2-9}
		  
		  & \bf{Scaling relations} & $\boldsymbol{N_{gal}=}223$
		  & $\boldsymbol{m_{\mathrm{F160W}}^{ref}=}17.34$
		  & $\boldsymbol{\alpha=}0.40$ & $\boldsymbol{\sigma_{LT}^{ref}=}235.7^{+21.8}_{-19.0}$ & $\boldsymbol{\beta_{cut}=}0.41$ & $\boldsymbol{r_{cut}^{ref}=}5.8^{+2.5}_{-1.8}$ & $\boldsymbol{\gamma=}0.20$\\
		  
		  \hline
		 
	\end{tabular}
	\\[6ex]

    \caption{
    {\it Top:} Input parameters of the reference model (\texttt{LM-model}) for the galaxy cluster \CL\ presented in this work. A single number is quoted for fixed parameters. When a flat prior on a free parameter is considered, the boundaries of the prior separated by the $\div$ symbol are reported. The number of galaxies optimized trough the scaling relations ($N_{gal}$), and the reference magnitude ($\boldsymbol{m_{\mathrm{F160W}}^{ref}}$) are also reported. 
    {\it Bottom:} Optimized values of the output parameters of the reference lens model. For each free parameter, we quote the median and the 16-th, and 84-th percentiles of the posterior distribution. 
    }    

	\label{table:inout_lensing}

\end{table*}

\begin{table}[h!]  
	\tiny
	\def\arraystretch{1.6}
	\centering    
	\begin{tabular}{|c|c|c|c|>{\centering\arraybackslash}m{9cm}|}
	   \hline
	   \multicolumn{4}{|c|}{\textbf{\normalsize Comparison between published lens models}}\\[2pt] 
	   \hline
	   \textbf{\normalsize Model} & \boldmath{\normalsize $N_\mathrm{images}$} & \boldmath{\normalsize $N_\mathrm{sources}$} & \boldmath{\normalsize $\Delta_{rms}\,\mathrm{[\arcsec]}$} \\[2pt] 
	   \hline
	   \textbf{This work} & \textbf{90} & \textbf{30} & \textbf{0.37} 
	   \cr
	   \hline
	   \citetalias{Richard_2021} & 83 & 29 & 0.67
	   \cr
	   \hline
	\end{tabular}
	\smallskip
    \caption{Comparison between our new lens model for \CL\ and other published models for the same cluster. \boldmath{\normalsize $N_\mathrm{images}$} is the number of multiple image used as model constraints, \boldmath{\normalsize $N_\mathrm{sources}$} is the number of background sources, and \boldmath{\normalsize $\Delta_{rms}\,\mathrm{[\arcsec]}$} is the total root-mean-square displacement between observed and predicted image positions (see \Eq\ref{eq.: rms_lt}).
    }
	\label{table:lens_models} 
\end{table}

\LT\ adopts a parametric approach, where the total mass distribution of a galaxy cluster is decomposed into the sum of several components. Extended HST imaging from the BUFFALO survey reveals several massive secondary structures in the outskirts, residing at distances between $\sim 600 - 775$ kpc from the BCG-N, and forming multiply imaged systems in the vicinity. The massive structures, confirmed to be at the cluster's redshift based on ancillary spectroscopy (see Section \ref{sec:dataMUSE}), can introduce a non-negligible perturbation in the positions of the multiple images in the core, and therefore impact the derived mass distribution \citep[e.g., ][]{Acebron2017}. In this work, we explore two different mass parametrizations of \CL. In the reference model, labeled \texttt{LM-model}, we model the cluster's environment as inferred from the extended BUFFALO imaging, while in the \texttt{WL-model}, we implement the results from previous WL studies \citep{Medezinski2016}.
The total mass distribution of the cluster is thus decomposed into the following mass contributions:

\begin{equation}
    \label{eq.: pot_dec}
    \phi_{tot}= \sum_{i=1}^{N_h}\phi_i^{halo}+\sum_{j=1}^{N_{bcg}}\phi_j^{BCG}+\sum_{k=1}^{N_g}\phi_k^{gal}+\sum_{l=1}^{N_s}\phi_l^{ENV}.
\end{equation}

\noindent The first component refers to the profiles used to parameterize the cluster-scale halos of the cluster (mainly made of dark matter). The second term corresponds to the mass contribution of the two brightest cluster galaxies (BCGs), BCG-N and BCG-S in Figure \ref{fig:RGB}, which are individually optimized in the lens model. The third sum describes the mass contribution of the cluster member galaxies (the subhalo component) to the total cluster mass, modeled within scaling relations. Finally, the fourth and last component models the contribution from structures in the cluster environment (where the parametrization of the two mass models differ). A detailed description of the modeling of each component is provided below. 

Both the cluster and subhalo mass component ($\phi_i^{halo}$, $\phi_j^{BCG}$ and $\phi_k^{gal}$) are described using dual pseudo-isothermal elliptical mass distributions \citep[dPIEs,][]{Limousin_lenstool, Eliasdottir_lenstool, Bergamini_2019}. This profile is characterized by seven free parameters: the position (x, y); the ellipticity (defined as $e=\frac{a^2-b^2}{a^2+b^2}$, where a and b are the semi-major and semi-minor axes of the ellipsoid, respectively); the position angle $\theta$ computed counterclockwise from the west direction; the central velocity dispersion $\sigma_{0}$ ;the core radius $r_{core}$; and the truncation radius $r_{cut}$. We note that, instead of using $\sigma_{0}$, \LT\ adopts a scaled version of this quantity, identified as $\sigma_{LT}$, such that $\sigma_{LT}=\sigma_{0} \sqrt{2/3}$.

The cluster-scale component of our new lens models ($\phi_i^{halo}$) is parametrized by two not truncated elliptical dPIEs, which are centered on the BCGs, denominated BCG-N and BCG-S in Figure \ref{fig:RGB}. The halos are left free to move within a small range around the BCG positions (see \T\,\ref{table:inout_lensing}).

Due to the presence of radial arcs in the vicinity of both BCGs (namely systems 2 and 4, see Figure \ref{fig:RGB} and Table \ref{tab:multiple_images}), and to improve their reconstruction, the parameters describing the mass contribution, and the ellipticity, of both BCGs are individually optimized (i.e., they are modeled outside of the scaling relations). This consists of 4 additional free parameters for each profile. 

Cluster member galaxies ($\phi_j^{gal}$) are described using singular circular dPIEs whose velocity dispersions, $\sigma^{gal}_{LT,i}$, and truncation radii, $r^{gal}_{cut,i}$, scale with the galaxy luminosity, $L_i$, according to the following scaling relations (which are used to sensibly reduce the number of free parameters of the lens model):

\begin{equation}
    \sigma^{gal}_{LT,i}= \sigma^{ref}_{LT} \left(  \frac{L_i}{L_{ref}} \right)^{\alpha},\quad
    r^{gal}_{cut,i}= r^{ref}_{cut} \left(  \frac{L_i}{L_{ref}} \right)^{\beta_{cut}}.
    \label{eq.: Scaling_relations}
\end{equation}
\medskip

\noindent The reference luminosity $L_{ref}=17.34$ corresponds to the BCG-N (see Figure \ref{fig:RGB}) magnitude in the HST F160W band. 
Following \cite{Bergamini_2021}, we fix $\alpha=0.40$ and $\beta_{cut}=0.41$. As described in \citet{Bergamini_2019}, these values are inferred from the measured inner stellar kinematics of 85 cluster member galaxies obtained by exploiting the MUSE data (see Section \ref{sec:cluster_members}).
A large, uniform, prior is assumed for $r^{ref}_{cut}$. 

As previously mentioned, the effect of the cluster environment is implemented in different ways in our SL modeling. In our reference model, \texttt{LM-model}, we include the mass contribution from the 3 brightest galaxies in the northern region of the cluster (called also external clumps), which are modeled as singular isothermal sphere profiles (SIS). Their positions are fixed to that of the light (see Table \ref{table:inout_lensing}), while their velocity dispersion values are optimized within flat large priors that account for the galaxy and large-scale DM distribution (adding 3 free parameters in total). \texttt{LM-model} has a total number of 25 free parameters related to the mass parametrization.
For the \texttt{WL-model}, the effect of the cluster environment is modeled based on the results from the WL analysis presented in \citet{Medezinski2016}, using deep imaging from Subaru/Suprime-Cam. Besides the main structure associated with the cluster's core, three additional sub-structures were detected with a high significance value of $\gtrsim 5\sigma$. These sub-structures \citep[labeled W, NE and NW in][and in Figure \ref{fig:Mass_image}]{Medezinski2016}, are modeled as SIS profiles. 
We adopt the best-fit positions and mass parameters from \citet{Medezinski2016} (Table 4), and their statistical uncertainties through large flat priors, in our lens model. Each halo adds thus 3 free parameters, the position, ($x$, $y$) and the velocity dispersion, $\sigma_0$. The \texttt{WL-model} has a total number of 31 free parameters related to the mass parametrization.

The priors assumed for the parameters of the mass profiles included in our reference lens model, \texttt{LM-model}, are summarized in the upper part of  \T\,\ref{table:inout_lensing}, while the optimized values are reported in the bottom.

\section{Results}
\label{sec:results}
The final $\Delta_{rms}$ value of our new reference lens model (\texttt{LM-model}) is $0.37\arcsec$, corresponding to an improvement of approximately a factor of $\sim2$ compared to the previous value ($0.67\arcsec$) obtained by \citetalias{Richard_2021}, where a smaller sample of multiple images was used to constrain the lens model. In Figure \ref{fig:rms}, we show the separations, along the $x$ and $y$ directions, between the observed and model predicted positions of the 90 multiple images (see also the red and green symbols in Figure \ref{fig:Magnification_image}). The figure shows that the model can reproduce accurately the observed position of the images in those cluster regions where the spatial density of image constraints is higher (more than five multiple images within $5\arcsec$). This result demonstrates that a large sample of secure multiple images, including resolved substructures within extended images (e.g., see the cutouts in Figure \ref{fig:RGB}), and distributed all around the cluster field, is an important ingredient to develop high-precision cluster strong lensing models \citep[see also][\ci{Bergamini et al. in prep.}]{Grillo2016, Bergamini_2020}.   

On the other hand, the optimization of the \texttt{WL-model}, which includes the three WL halos, leads to a $\Delta_{rms}$ value of $0.44\arcsec$, therefore significantly larger than our reference model. This underscores the better ability of the light-traces-mass approach in reproducing the mass distribution of the external region of the cluster.  

\begin{figure*}[h]
	\centering
	\includegraphics[width=\linewidth]{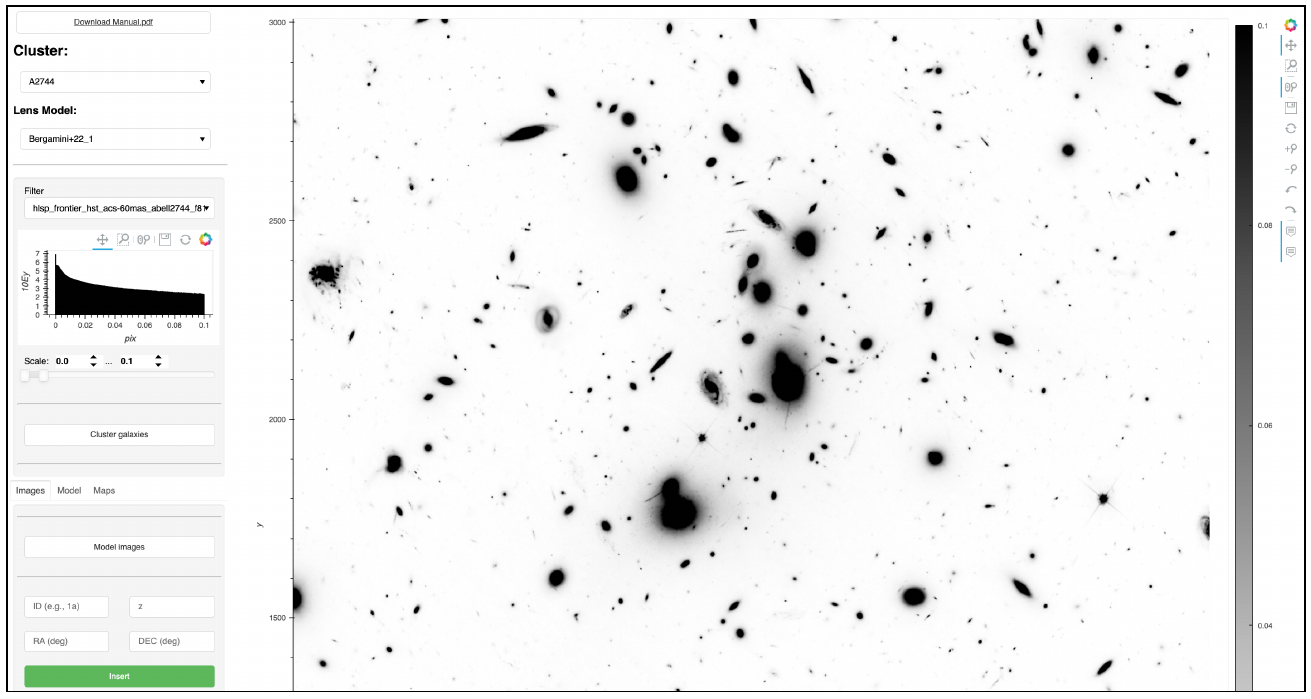}
	\caption{The graphical interface of our new Strong Lensing Online Tool (\SLOT) allows for a full and easy access to high-level products, including the statistical uncertainties, of our lens models. The link will be available upon publication.
	}
	\label{SLOT}
\end{figure*}

In Figure \ref{fig:Mass_image}, we show the resulting total projected mass distribution of \CL\ from our reference best-fit lens model superimposed to the HST/F814W image. We note that our lens model can be considered accurate up to $50$-$60\arcsec$ from the BCG-N (this is the largest distance at which secure multiple images have been identified). Model predictions outside that region are extrapolations and can, therefore, be prone to systematic errors.

Figure \ref{fig:mass_profile} shows that the total projected cluster mass distribution within 300\,kpc from the BCG-N (in black), and its different components (as described in Equation \ref{eq.: pot_dec}). 
Due to the degeneracies between model parameters, a clear separation between the mass contributions from the BCGs and the cluster halo is not possible. The contribution of the member galaxies to the cluster total mass decreases as a function of the distance from the cluster center.
At radii larger than 100\,kpc, the mass contribution of the external clumps dominates over that of the subhalo. At a radius of 200\,kpc, the total mass of the cluster is $M_{\rm tot}=(1.77 ~\pm 0.07)\times10^{14}\,\mathrm{M_{\odot}}$, including both the statistical and systematic uncertainties (see Figure \ref{fig:mass_profile}).  

In \Fig\,\ref{fig:green_plot}, we show the $\sigma$-luminosity relation. For a proper comparison of the lensing velocity dispersion of the galaxies with the observed ones, the values are corrected for the spectroscopic aperture (0.8\arcsec), as detailed in \citep[see][]{Bergamini_2019}). The cluster member scaling relations inferred from the lens models are in excellent agreement with the kinematic results, although no prior on the $\sigma_{LT}^{ref}$ value is assumed. We have checked that by imposing a prior on that value, we find no significant difference in the $\Delta_{rms}$ value (i.e., $\sim0.03\arcsec$ higher).

In Figure \ref{fig:Magnification_image}, we show the absolute magnification map computed for a background source at $z=3$. While the magnification values are significantly different from one at large distances from the cluster center, we caution that these values are extrapolations in regions that lack strong lensing constraints. 
Relatively large magnification values are also supported by the presence of several strong lensing features that are visible at distances between $\sim 600 - 775$ kpc from the BCG-N, revealed in the BUFFALO imaging. The formation of these distant gravitational arcs is attributed to the presence of secondary cluster clumps surrounding \CL. Their mass is expected to be comparable to those of the main cluster (see \T\,\ref{table:inout_lensing}).

\section{\SLOT: Strong Lensing Online Tool}
\label{sec:SLOT}
The \CL\ lens model presented in this work will be made publicly available with the publication of this paper. In order to allow interested users to access and exploit the model results, we have developed a graphical interface, pictured in Figure \ref{SLOT}. The new Strong Lensing Online Tool (\SLOT) will allow astronomers to take full advantage of the predictive and statistical results of our high-precision strong lensing model for their research, both for studies on cluster lenses and high-redshift sources. For example, \SLOT\ can be used to: compute magnification values, with a careful estimation of the associated statistical errors, for all the sources in the field of \CL; obtain the predicted positions of the counter-images of every source defined by the user; derive and download maps of projected cluster total mass, deflection angle, magnification, etc. A description of the functionality of \SLOT\ and the user manual can be found in the
{\SLOT\ webpage} (the link will be available upon publication).  

\section{Conclusions}
\label{sec:conclusions}

We have presented a new high-precision strong lensing model for the galaxy cluster \CL. A careful inspection of the HST and MUSE data has allowed us to create a secure image data-set counting 90 multiple images (from 30 background sources) whose point-like positions are used to constrain the lens model. This is currently the largest multiple image catalog compiled for this cluster.
Our lens model also includes the information coming from the measured stellar kinematics of 85 member galaxies that is used to accurately characterize the subhalo component of the cluster. This component counts a total of 225 galaxies (including the two cluster BCGs), 202 of which are spectroscopically confirmed cluster members. The gravitational lens \CL\ is strongly affected by the presence of massive structures in the North-West region of the cluster,
at distances between $\sim 600 - 775$ kpc from the BCG-N. The contribution of these outer massive structures is  taken into account by using three additional SIS profiles in our reference lens model fixed on the positions of the three brightest galaxies in that area. 

The final $\Delta_{rms}$ value of our reference model, \texttt{LM-model}, is equal to $0.37\arcsec$ representing a significant step forward with respect to previous SL models for the same cluster, and with the same data-set (e.g., \citetalias{Richard_2021}). The inclusion of the sub-structures within several extended sources as model constraints, including the radial arcs, has allowed us to accurately characterize the inner total mass distribution of the cluster and the position of the cluster critical lines.

In addition, we have found that accounting for the mass distribution of the cluster outside the core, particularly using the light of three prominent cluster galaxies as total mass tracers, is a key ingredient to an accurate prediction of the positions of the multiple images.

Finally, we have presented a new publicly available graphical interface, called Strong Lensing Online Tool (\texttt{SLOT}). Interested users, even non-lensing experts, can exploit the predictive power and the full statistical information of the lens model presented in this work through a user-friendly graphical interface.

We plan to apply our new high-precision strong lensing model for the first analysis of the GLASS-JWST-ERS observations, specifically to make use of magnification values and uncertainties for high-$z$ lensed sources. Clearly, JWST imaging and spectroscopic data will reveal a much larger number of multiply imaged sources with resolved substructures, over a more extended redshift range, which we will employ to improve the lens model presented in this work.

\begin{acknowledgements}
Based on observations collected at the European Southern Observatory for Astronomical research in the Southern Hemisphere under ESO programmes with IDs 094.A-0115 (PI: Richard).
We acknowledge financial support through grants PRIN-MIUR 2015W7KAWC, 2017WSCC32, and 2020SKSTHZ. AA has received funding from the European Union’s Horizon 2020 research and innovation programme under the Marie Skłodowska-Curie grant agreement No 101024195 — ROSEAU. GBC thanks the Max Planck Society for support through the Max Planck Research Group for S. H. Suyu and the academic support from the German Centre for Cosmological Lensing. MM acknowledges support from the Italian Space Agency (ASI) through contract ``Euclid - Phase D". We acknowledge funding from the INAF ``main-stream'' grants 1.05.01.86.20 and  1.05.01.86.31.
\end{acknowledgements}

%
%

\bibliographystyle{aa}
\bibliography{bibliography}

\begin{appendix}

\section{Multiple images}
We present the catalog of the 90 secure multiple images (from 30 background sources), that are included as constraints in our SL model, and that represent the largest secure sample in the \CL\ cluster field. All systems are discussed, and compared to the catalog presented in \citetalias{Richard_2021}, in Section \ref{sec:multiple_images}.

\begin{table}[h]
\caption{Catalog of the spectroscopic multiple images included in the SL modeling of \CL.
} 
\label{tab:multiple_images}
\centering
\begin{tabular}{cccccc}
\hline\hline
ID & R.A. & Decl & $\rm z_{spec}$ & QF & QP\\ 
 & deg & deg &  &  &  \\ 
\hline
\vspace{-0.2cm}\\
1.1a & 3.597561 & -30.403925 & 1.688100 & 3 & 1 \\ 
1.1b & 3.595963 & -30.406808 & 1.688100 & 2 & 1 \\ 
1.1c & 3.586226 & -30.409986 & 1.688100 & 3 & 1 \\ 
1.2a & 3.597072 & -30.404723 & 1.688100 & 3 & 1 \\ 
1.2b & 3.596395 & -30.406143 & 1.688100 & 3 & 1 \\ 
1.2c & 3.585748 & -30.410100 & 1.688100 & 3 & 1 \\ 
1.3a & 3.597756 & -30.403530 & 1.688100 & 3 & 2 \\ 
1.3b & 3.595528 & -30.407199 & 1.688100 & 3 & 2 \\ 
1.3c & 3.586459 & -30.409871 & 1.688100 & 3 & 2 \\ 
1.4a & 3.598082 & -30.403980 & 1.688100 & 1 & 1 \\ 
1.4b & 3.595722 & -30.407546 & 1.688100 & 1 & 1 \\ 
1.4c & 3.587383 & -30.410152 & 1.688100 & 1 & 1 \\ 
2.1a & 3.583265 & -30.403339 & 1.886800 & 3 & 1 \\ 
2.1b & 3.597289 & -30.396712 & 1.886800 & 3 & 1 \\ 
2.1c & 3.585369 & -30.399878 & 1.886800 & 3 & 1 \\ 
2.1d & 3.586412 & -30.402127 & 1.886800 & 3 & 1 \\ 
2.2a & 3.583029 & -30.403189 & 1.886800 & 3 & 1 \\ 
2.2b & 3.597138 & -30.396639 & 1.886800 & 3 & 1 \\ 
2.2c & 3.585134 & -30.399668 & 1.886800 & 3 & 1 \\ 
2.2d & 3.586438 & -30.401870 & 1.886800 & 3 & 1 \\ 
2.3a & 3.582994 & -30.403050 & 1.886800 & 3 & 1 \\ 
2.3b & 3.597095 & -30.396580 & 1.886800 & 3 & 1 \\ 
2.3c & 3.585017 & -30.399625 & 1.886800 & 3 & 1 \\ 
2.3d & 3.586394 & -30.401765 & 1.886800 & 3 & 1 \\ 
2.4a & 3.582918 & -30.402930 & 1.886800 & 3 & 1 \\ 
2.4b & 3.597052 & -30.396527 & 1.886800 & 3 & 1 \\ 
2.4c & 3.584915 & -30.399580 & 1.886800 & 3 & 1 \\ 
2.4d & 3.586353 & -30.401632 & 1.886800 & 3 & 1 \\ 
2.5a & 3.582828 & -30.402793 & 1.886800 & 3 & 2 \\ 
2.5b & 3.596990 & -30.396469 & 1.886800 & 3 & 2 \\ 
2.5c & 3.584819 & -30.399517 & 1.886800 & 3 & 2 \\ 
2.5d & 3.586322 & -30.401490 & 1.886800 & 3 & 1 \\ 
2.6a & 3.582772 & -30.402683 & 1.886800 & 3 & 2 \\ 
2.6b & 3.596942 & -30.396421 & 1.886800 & 3 & 2 \\ 
2.6c & 3.584722 & -30.399463 & 1.886800 & 3 & 2 \\ 
2.6d & 3.584719 & -30.399462 & 1.886800 & 3 & 2 \\ 
2.7a & 3.582530 & -30.402313 & 1.886800 & 3 & 2 \\ 
2.7b & 3.596734 & -30.396291 & 1.886800 & 3 & 2 \\ 
2.7c & 3.584473 & -30.399289 & 1.886800 & 3 & 2 \\ 
2.7d & 3.586236 & -30.400869 & 1.886800 & 3 & 2 \\ 
3.1a & 3.588807 & -30.393797 & 3.980500 & 3 & 1 \\ 
3.1b & 3.589375 & -30.393857 & 3.980500 & 3 & 1 \\ 
3.2a & 3.589222 & -30.393844 & 3.980500 & 3 & 1 \\ 
3.2b & 3.588965 & -30.393815 & 3.980500 & 3 & 1 \\ 
3.3a & 3.589487 & -30.393869 & 3.980500 & 3 & 1 \\ 
\hline  
\end{tabular}
\end{table}

\begin{table}[h]
\vspace{4.08cm}
\centering
\begin{tabular}{cccccc}
\hline\hline
ID & R.A. & Decl & $\rm z_{spec}$ & QF & QP\\ 
 & deg & deg &  &  &  \\ 
\hline
\vspace{-0.2cm}\\
3.3b & 3.588639 & -30.393785 & 3.980500 & 3 & 1 \\ 
4.1a & 3.592126 & -30.402660 & 3.577000 & 3 & 1 \\ 
4.1b & 3.595674 & -30.401633 & 3.577000 & 3 & 1 \\ 
4.1c & 3.580452 & -30.408946 & 3.577000 & 3 & 2 \\ 
4.1e & 3.593651 & -30.405117 & 3.577000 & 3 & 3 \\ 
4.2a & 3.592098 & -30.402527 & 3.577000 & 3 & 1 \\ 
4.2b & 3.595567 & -30.401518 & 3.577000 & 3 & 1 \\ 
6.1a & 3.598541 & -30.401792 & 2.017100 & 3 & 1 \\ 
6.1b & 3.594058 & -30.407999 & 2.017100 & 3 & 1 \\ 
6.1c & 3.586433 & -30.409363 & 2.017100 & 3 & 1 \\ 
8.1a & 3.589712 & -30.394335 & 3.976400 & 2 & 2 \\ 
8.1b & 3.588831 & -30.394202 & 3.976800 & 2 & 2 \\ 
18.1a & 3.576124 & -30.404472 & 5.662400 & 3 & 1 \\ 
18.1b & 3.588377 & -30.395630 & 5.662400 & 3 & 1 \\ 
18.1c & 3.590730 & -30.395543 & 5.662400 & 3 & 1 \\ 
22.1a & 3.587924 & -30.411608 & 5.284400 & 3 & 2 \\ 
22.1b & 3.600048 & -30.404415 & 5.284400 & 3 & 2 \\ 
22.1c & 3.596592 & -30.408989 & 5.283700 & 3 & 2 \\ 
26.1a & 3.593898 & -30.409724 & 3.054200 & 3 & 1 \\ 
26.1b & 3.590353 & -30.410575 & 3.054200 & 3 & 1 \\ 
26.1c & 3.600112 & -30.402939 & 3.054200 & 9 & 2 \\ 
26.2a & 3.593993 & -30.409699 & 3.054200 & 3 & 1 \\ 
26.2b & 3.590272 & -30.410610 & 3.054200 & 3 & 1 \\ 
26.3a & 3.594031 & -30.409604 & 3.054200 & 2 & 3 \\ 
26.3b & 3.589969 & -30.410593 & 3.054200 & 2 & 3 \\ 
33.1a & 3.584712 & -30.403146 & 5.725600 & 3 & 1 \\ 
33.1b & 3.584397 & -30.403393 & 5.725600 & 3 & 1 \\ 
34.1a & 3.593428 & -30.410834 & 3.784000 & 2 & 1 \\ 
34.1b & 3.593812 & -30.410714 & 3.784000 & 3 & 1 \\ 
34.1c & 3.600711 & -30.404593 & 3.787300 & 2 & 1 \\ 
42.1a & 3.597313 & -30.400605 & 3.691900 & 3 & 1 \\ 
42.1b & 3.590956 & -30.403252 & 3.691900 & 3 & 1 \\ 
42.1c & 3.581590 & -30.408631 & 3.691900 & 3 & 1 \\ 
42.1d & 3.594245 & -30.406388 & 3.691900 & 3 & 1 \\ 
42.1e & 3.592415 & -30.405197 & 3.691900 & 1 & 3 \\ 
61.1a & 3.595522 & -30.403485 & 2.950900 & 1 & 1 \\ 
61.1b & 3.595138 & -30.404471 & 2.950900 & 3 & 1 \\ 
62.1a & 3.591326 & -30.398643 & 4.193600 & 3 & 3 \\ 
62.1b & 3.590582 & -30.398918 & 4.193600 & 3 & 3 \\ 
63.1a & 3.582214 & -30.407142 & 5.661500 & 3 & 1 \\ 
63.1b & 3.592836 & -30.407032 & 5.661500 & 3 & 2 \\ 
63.1c & 3.589153 & -30.403427 & 5.661500 & 3 & 1 \\ 
63.1d & 3.598830 & -30.398273 & 5.661500 & 3 & 1 \\ 
64.1a & 3.581203 & -30.398734 & 3.408600 & 3 & 1 \\ 
64.1c & 3.596420 & -30.394264 & 3.408600 & 3 & 1 \\ 
\hline  
\end{tabular}
\end{table}

\clearpage

\section{Cluster members}
\label{sec:cluster_members_appendix}
We present the catalog of the 225 cluster member galaxies that are included in our SL model, 202 of which are spectroscopically selected (based on the MUSE and ancillary spectroscopy, presented in Section \ref{sec:dataMUSE}) and 23 are identified based on HST multi-band photometry through a CNN technique (see Section \ref{sec:cluster_members}). 

\begin{table}[h!]
\caption{Catalog of the spectroscopic (top) and photometric (bottom) cluster members included in the SL modeling of \CL.
} 
\label{tab:cluster_members}
\centering
\begin{tabular}{ccccc}
\hline\hline
ID & R.A. & Decl & $m_{F160W}$ & $\rm z_{spec}$\\ 
 & deg & deg &  &  \\ 
\hline
\vspace{-0.2cm}\\
36034 & 3.592037 & -30.405741 & 17.30 & 0.3185\tablefootmark{a} \\ 
37824 & 3.586257 & -30.400172 & 17.34 & 0.2997\tablefootmark{a} \\ 
835 & 3.589290 & -30.369074 & 16.88 & 0.3002\tablefootmark{b} \\ 
40689 & 3.594796 & -30.391654 & 17.58 & 0.3006\tablefootmark{a} \\ 
736 & 3.575038 & -30.428344 & 17.69 & 0.3170\tablefootmark{b} \\ 
938 & 3.610666 & -30.395618 & 17.70 & 0.3033\tablefootmark{b} \\ 
814 & 3.587469 & -30.371246 & 17.76 & 0.3038\tablefootmark{b} \\ 
947 & 3.612408 & -30.409245 & 17.84 & 0.3031\tablefootmark{b} \\ 
34423 & 3.579662 & -30.409189 & 17.84 & 0.3027\tablefootmark{a} \\ 
40059 & 3.585386 & -30.394279 & 17.89 & 0.3202\tablefootmark{a} \\ 
783 & 3.583084 & -30.433553 & 18.00 & 0.2927\tablefootmark{b} \\ 
20227 & 3.609542 & -30.382110 & 18.03 & 0.3200\tablefootmark{c} \\ 
39382 & 3.587646 & -30.396426 & 18.07 & 0.3031\tablefootmark{a} \\ 
36210 & 3.592510 & -30.404611 & 18.24 & 0.3150\tablefootmark{a} \\ 
809 & 3.586500 & -30.367380 & 18.29 & 0.3007\tablefootmark{b} \\ 
697 & 3.566354 & -30.388260 & 18.40 & 0.3025\tablefootmark{e} \\ 
36527 & 3.578511 & -30.403375 & 18.41 & 0.3155\tablefootmark{a} \\ 
38067 & 3.574900 & -30.398381 & 18.46 & 0.3175\tablefootmark{a} \\ 
740 & 3.575081 & -30.377074 & 18.49 & 0.3142\tablefootmark{b} \\ 
730 & 3.573503 & -30.422861 & 18.50 & 0.3188\tablefootmark{b} \\ 
37947 & 3.592998 & -30.399330 & 18.57 & 0.3092\tablefootmark{a} \\ 
834 & 3.589103 & -30.419803 & 18.59 & 0.3044\tablefootmark{b} \\ 
894 & 3.602057 & -30.377689 & 18.60 & 0.3011\tablefootmark{b} \\ 
41259 & 3.593289 & -30.384378 & 18.60 & 0.2964\tablefootmark{a} \\ 
40592 & 3.589220 & -30.389839 & 18.64 & 0.3151\tablefootmark{a} \\ 
41644 & 3.570173 & -30.386449 & 18.66 & 0.2969\tablefootmark{a} \\ 
37954 & 3.586559 & -30.399391 & 18.70 & 0.3229\tablefootmark{a} \\ 
35061 & 3.573945 & -30.408829 & 18.71 & 0.3135\tablefootmark{a} \\ 
39428 & 3.588152 & -30.395075 & 18.71 & 0.3002\tablefootmark{a} \\ 
41950 & 3.589184 & -30.387396 & 18.72 & 0.3169\tablefootmark{a} \\ 
38907 & 3.585204 & -30.394649 & 18.81 & 0.3009\tablefootmark{a} \\ 
642 & 3.556336 & -30.387018 & 18.88 & 0.3115\tablefootmark{b} \\ 
38117 & 3.582159 & -30.398571 & 18.89 & 0.2986\tablefootmark{a} \\ 
39072 & 3.598969 & -30.397533 & 18.90 & 0.3162\tablefootmark{a} \\ 
41856 & 3.585314 & -30.387545 & 18.91 & 0.3008\tablefootmark{a} \\ 
804 & 3.585629 & -30.366902 & 18.98 & 0.2998\tablefootmark{b} \\ 
37344 & 3.604341 & -30.400124 & 18.98 & 0.3190\tablefootmark{a} \\ 
41440 & 3.605429 & -30.384843 & 18.99 & 0.3112\tablefootmark{a} \\ 
38010 & 3.588385 & -30.398355 & 19.01 & 0.3173\tablefootmark{a} \\ 
678 & 3.562510 & -30.402406 & 19.05 & 0.3025\tablefootmark{b} \\ 
38930 & 3.588680 & -30.396077 & 19.07 & 0.3020\tablefootmark{a} \\ 
40243 & 3.580953 & -30.390808 & 19.15 & 0.2931\tablefootmark{a} \\ 
32284 & 3.602650 & -30.416956 & 19.16 & 0.3132\tablefootmark{a}\\  
\hline
\end{tabular}
\tablefoot{
\tablefoottext{a}{MUSE measurement from this work}\\
\tablefoottext{b}{\citet{Owers2011}}\\
\tablefoottext{c}{\citet{Treu2015, Schmidt_2014}}\\
\tablefoottext{d}{\citet{Braglia2009}}\\
\tablefoottext{e}{\citet{Couch1987}}
}
\end{table}

\begin{table}[h]

\label{tab:cluster_members}
\centering
\begin{tabular}{ccccc}
\hline\hline
ID & R.A. & Decl & $m_{F160W}$ & $\rm z_{spec}$\\ 
 & deg & deg &  &  \\ 
\hline
\vspace{-0.2cm}\\
690 & 3.565544 & -30.387093 & 19.16 & 0.2991\tablefootmark{b} \\
40478 & 3.571507 & -30.390436 & 19.18 & 0.2965\tablefootmark{a} \\

20018 & 3.581566 & -30.376517 & 19.22 & 0.3130\tablefootmark{c} \\ 
902 & 3.604284 & -30.414554 & 19.25 & 0.3061\tablefootmark{b} \\ 

966 & 3.618050 & -30.403670 & 19.26 & 0.3115\tablefootmark{b} \\
42443 & 3.594708 & -30.389115 & 19.26 & 0.3035\tablefootmark{a} \\ 
40314 & 3.590342 & -30.390939 & 19.26 & 0.2972\tablefootmark{a} \\ 
720 & 3.571365 & -30.422840 & 19.29 & 0.3030\tablefootmark{b} \\ 
41363 & 3.588146 & -30.385006 & 19.30 & 0.2977\tablefootmark{a} \\ 
692 & 3.565347 & -30.382948 & 19.38 & 0.3030\tablefootmark{b} \\ 
768 & 3.580711 & -30.418875 & 19.39 & 0.2934\tablefootmark{b} \\ 
956 & 3.615142 & -30.383729 & 19.41 & 0.3061\tablefootmark{b} \\ 
38729 & 3.578864 & -30.397111 & 19.42 & 0.3190\tablefootmark{a} \\ 
973 & 3.618739 & -30.392932 & 19.42 & 0.3015\tablefootmark{b} \\ 
36892 & 3.587938 & -30.400852 & 19.44 & 0.3150\tablefootmark{a} \\ 
863 & 3.593561 & -30.426049 & 19.46 & 0.2968\tablefootmark{b} \\ 
35339 & 3.595907 & -30.406213 & 19.49 & 0.3161\tablefootmark{a} \\ 
950 & 3.613184 & -30.389698 & 19.49 & 0.3026\tablefootmark{b} \\ 
20132 & 3.595399 & -30.380404 & 19.50 & 0.3200\tablefootmark{c} \\ 
921 & 3.607012 & -30.403478 & 19.51 & 0.2966\tablefootmark{b} \\ 
655 & 3.559037 & -30.410659 & 19.51 & 0.2984\tablefootmark{b} \\ 
32547 & 3.601355 & -30.415365 & 19.51 & 0.3197\tablefootmark{a} \\ 
37068 & 3.605266 & -30.400808 & 19.52 & 0.3197\tablefootmark{a} \\ 
888 & 3.600962 & -30.417843 & 19.59 & 0.3042\tablefootmark{a} \\ 
41418 & 3.592547 & -30.385314 & 19.62 & 0.3164\tablefootmark{a} \\ 
931 & 3.609576 & -30.378771 & 19.62 & 0.3009\tablefootmark{b} \\ 
39503 & 3.581389 & -30.393932 & 19.63 & 0.2998\tablefootmark{a} \\ 
634 & 3.555379 & -30.384632 & 19.66 & 0.3007\tablefootmark{b} \\ 
34556 & 3.591721 & -30.407807 & 19.71 & 0.3194\tablefootmark{a} \\ 
816 & 3.587541 & -30.373945 & 19.71 & 0.2963\tablefootmark{b} \\ 
961 & 3.616679 & -30.402709 & 19.72 & 0.2943\tablefootmark{b} \\ 
41303 & 3.583714 & -30.384680 & 19.75 & 0.3029\tablefootmark{a} \\ 
33910 & 3.589134 & -30.409573 & 19.75 & 0.3173\tablefootmark{a} \\ 
42149 & 3.598764 & -30.388018 & 19.76 & 0.3027\tablefootmark{a} \\ 
33540 & 3.588817 & -30.410722 & 19.77 & 0.3223\tablefootmark{a} \\ 
787 & 3.583285 & -30.432301 & 19.78 & 0.2942\tablefootmark{b} \\ 
39646 & 3.578948 & -30.394119 & 19.80 & 0.3191\tablefootmark{a} \\ 
40884 & 3.590278 & -30.382698 & 19.86 & 0.3019\tablefootmark{a} \\ 
37229 & 3.594463 & -30.400350 & 19.89 & 0.3036\tablefootmark{a} \\ 
33328 & 3.569589 & -30.412164 & 19.95 & 0.2990\tablefootmark{a} \\ 
40270 & 3.594239 & -30.390462 & 19.97 & 0.3163\tablefootmark{a} \\ 
42269 & 3.595506 & -30.388688 & 19.97 & 0.3032\tablefootmark{a} \\ 
13996 & 3.573450 & -30.377932 & 20.00 & 0.3184\tablefootmark{d} \\ 
39710 & 3.584986 & -30.392877 & 20.01 & 0.2954\tablefootmark{a} \\ 
35693 & 3.587039 & -30.404948 & 20.04 & 0.2987\tablefootmark{a} \\ 
40551 & 3.589520 & -30.389499 & 20.08 & 0.2939\tablefootmark{a} \\ 
21367 & 3.597859 & -30.405556 & 20.12 & 0.3208\tablefootmark{a} \\ 
40832 & 3.588038 & -30.382557 & 20.17 & 0.3116\tablefootmark{a} \\ 
39876 & 3.580373 & -30.392204 & 20.21 & 0.2935\tablefootmark{a} \\ 
33699 & 3.582507 & -30.409986 & 20.24 & 0.3188\tablefootmark{a} \\ 
38275 & 3.585521 & -30.397156 & 20.24 & 0.3124\tablefootmark{a} \\ 
41655 & 3.573735 & -30.385976 & 20.26 & 0.2965\tablefootmark{a} \\ 
41651 & 3.573383 & -30.386313 & 20.34 & 0.3079\tablefootmark{a} \\ 
37230 & 3.583991 & -30.399260 & 20.34 & 0.3200\tablefootmark{a} \\ 
40802 & 3.570861 & -30.382004 & 20.37 & 0.3066\tablefootmark{a} \\ 
36220 & 3.605275 & -30.402932 & 20.39 & 0.3161\tablefootmark{a} \\ 
39283 & 3.600830 & -30.394896 & 20.39 & 0.3061\tablefootmark{a} \\ 
713 & 3.569303 & -30.384236 & 20.42 & 0.2961\tablefootmark{b} \\ 
20089 & 3.597633 & -30.379200 & 20.51 & 0.2800\tablefootmark{c} \\ 
32768 & 3.585709 & -30.413971 & 20.59 & 0.3023\tablefootmark{a} \\ 
\hline
\end{tabular}
\end{table}

\begin{table}[h]

\label{tab:cluster_members}
\centering
\begin{tabular}{ccccc}
\hline\hline
ID & R.A. & Decl & $m_{F160W}$ & $\rm z_{spec}$\\ 
 & deg & deg &  &  \\ 
\hline
\vspace{-0.2cm}\\

42195 & 3.578348 & -30.387100 & 20.61 & 0.3076\tablefootmark{a} \\ 
13311 & 3.560070 & -30.389342 & 20.61 & 0.2999\tablefootmark{d} \\ 
33870 & 3.595124 & -30.409366 & 20.63 & 0.3199\tablefootmark{a} \\ 
35908 & 3.585035 & -30.403315 & 20.63 & 0.3037\tablefootmark{a} \\ 
37609 & 3.578591 & -30.399109 & 20.68 & 0.3054\tablefootmark{a} \\ 
36043 & 3.584377 & -30.402887 & 20.83 & 0.3159\tablefootmark{a} \\ 
41908 & 3.604401 & -30.384960 & 20.84 & 0.2962\tablefootmark{a} \\ 
34439 & 3.590280 & -30.407401 & 20.92 & 0.3182\tablefootmark{a} \\ 
40032 & 3.593885 & -30.390837 & 20.96 & 0.2970\tablefootmark{a} \\ 
36339 & 3.588145 & -30.401980 & 20.97 & 0.2986\tablefootmark{a} \\ 
32088 & 3.603422 & -30.416769 & 20.99 & 0.3179\tablefootmark{a} \\ 
38143 & 3.602200 & -30.396988 & 21.01 & 0.3031\tablefootmark{a} \\ 
44545 & 3.578469 & -30.381318 & 21.04 & 0.3179\tablefootmark{a} \\ 
36953 & 3.572810 & -30.400534 & 21.05 & 0.3143\tablefootmark{a} \\ 
37825 & 3.602715 & -30.397571 & 21.08 & 0.2990\tablefootmark{a} \\ 
40428 & 3.578347 & -30.389464 & 21.12 & 0.2947\tablefootmark{a} \\ 
20064 & 3.577544 & -30.378870 & 21.26 & 0.3080\tablefootmark{c} \\ 
36298 & 3.594832 & -30.402130 & 21.30 & 0.3162\tablefootmark{a} \\ 
35576 & 3.587971 & -30.404253 & 21.31 & 0.2998\tablefootmark{a} \\ 
40239 & 3.593166 & -30.390349 & 21.33 & 0.3013\tablefootmark{a} \\ 
36849 & 3.596757 & -30.400513 & 21.33 & 0.3167\tablefootmark{a} \\ 
34538 & 3.573400 & -30.407461 & 21.46 & 0.3141\tablefootmark{a} \\ 
36843 & 3.579074 & -30.400089 & 21.50 & 0.3056\tablefootmark{a} \\ 
40703 & 3.596256 & -30.388517 & 21.54 & 0.2964\tablefootmark{a} \\ 
33671 & 3.583831 & -30.409506 & 21.56 & 0.3023\tablefootmark{a} \\ 
33410 & 3.589721 & -30.410222 & 21.57 & 0.3154\tablefootmark{a} \\ 
40708 & 3.571061 & -30.388157 & 21.61 & 0.3032\tablefootmark{a} \\ 
38252 & 3.592874 & -30.396356 & 21.61 & 0.2975\tablefootmark{a} \\ 
37214 & 3.585389 & -30.399007 & 21.67 & 0.3006\tablefootmark{8} \\ 
36163 & 3.588499 & -30.402102 & 21.70 & 0.2992\tablefootmark{a} \\ 
39727 & 3.584373 & -30.391755 & 21.75 & 0.3200\tablefootmark{a} \\ 
41636 & 3.573693 & -30.385685 & 21.79 & 0.2971\tablefootmark{a} \\ 
37542 & 3.584458 & -30.398342 & 22.02 & 0.3235\tablefootmark{a} \\ 
38175 & 3.583312 & -30.396542 & 22.05 & 0.3073\tablefootmark{a} \\ 
35190 & 3.601244 & -30.404885 & 22.11 & 0.3049\tablefootmark{a} \\ 
38267 & 3.597298 & -30.396046 & 22.15 & 0.3189\tablefootmark{a} \\ 
33803 & 3.593740 & -30.409224 & 22.18 & 0.3033\tablefootmark{a} \\ 
37367 & 3.601857 & -30.398661 & 22.22 & 0.3144\tablefootmark{a} \\ 
42079 & 3.593478 & -30.387595 & 22.24 & 0.2962\tablefootmark{a} \\ 
8024000 & 3.567362 & -30.400881 & 22.30 & 0.3023\tablefootmark{a} \\ 
34705 & 3.594676 & -30.405899 & 22.31 & 0.3072\tablefootmark{a} \\ 
41265 & 3.578737 & -30.384233 & 22.33 & 0.3168\tablefootmark{a} \\ 
37199 & 3.603831 & -30.399122 & 22.37 & 0.3026\tablefootmark{a} \\ 
41937 & 3.573015 & -30.387260 & 22.43 & 0.3219\tablefootmark{a} \\ 
35514 & 3.574778 & -30.403672 & 22.43 & 0.3039\tablefootmark{a} \\ 
32680 & 3.592097 & -30.413109 & 22.45 & 0.3067\tablefootmark{a} \\ 
36814 & 3.593414 & -30.400024 & 22.49 & 0.3040\tablefootmark{a} \\ 
36982 & 3.582898 & -30.399701 & 22.49 & 0.2914\tablefootmark{a} \\ 
40944 & 3.593025 & -30.382970 & 22.50 & 0.3005\tablefootmark{a} \\ 
37231 & 3.581618 & -30.399105 & 22.51 & 0.3049\tablefootmark{a} \\ 
10440000 & 3.569048 & -30.394963 & 22.56 & 0.3077\tablefootmark{a} \\ 
33503 & 3.597475 & -30.409964 & 22.56 & 0.3051\tablefootmark{a} \\ 
10657000 & 3.569981 & -30.394634 & 22.61 & 0.3011\tablefootmark{a} \\ 
41388 & 3.598749 & -30.384458 & 22.61 & 0.3089\tablefootmark{a} \\ 
36872 & 3.595752 & -30.399971 & 22.62 & 0.3160\tablefootmark{a} \\ 
34828 & 3.592717 & -30.406107 & 22.65 & 0.3187\tablefootmark{8} \\ 
38459 & 3.572068 & -30.395969 & 22.66 & 0.2990\tablefootmark{a} \\ 
39956 & 3.584993 & -30.390870 & 22.67 & 0.3161\tablefootmark{a} \\ 
33933 & 3.585939 & -30.408434 & 22.71 & 0.3080\tablefootmark{a} \\ 
41467 & 3.586832 & -30.384627 & 22.74 & 0.3085\tablefootmark{a} \\ 
41842 & 3.591221 & -30.386717 & 22.76 & 0.3132\tablefootmark{a} \\

\hline
\end{tabular}

\end{table}

\begin{table}[h]

\label{tab:cluster_members}
\centering
\begin{tabular}{ccccc}
\hline\hline
ID & R.A. & Decl & $m_{F160W}$ & $\rm z_{spec}$\\ 
 & deg & deg &  &  \\ 
\hline
\vspace{-0.2cm}\\

38253 & 3.584471 & -30.396056 & 22.78 & 0.3224\tablefootmark{a} \\ 
41531 & 3.589828 & -30.385639 & 22.85 & 0.3174\tablefootmark{a} \\ 
4206000 & 3.600810 & -30.409530 & 22.87 & 0.3068\tablefootmark{a} \\ 
36346 & 3.588851 & -30.401771 & 22.89 & 0.3059\tablefootmark{a} \\ 
41930 & 3.608089 & -30.387068 & 22.90 & 0.2962\tablefootmark{a} \\ 
33753 & 3.567161 & -30.408755 & 22.92 & 0.3170\tablefootmark{a} \\ 
33911 & 3.577555 & -30.408196 & 22.94 & 0.3068\tablefootmark{a} \\ 
39609 & 3.605171 & -30.392616 & 22.98 & 0.3217\tablefootmark{a} \\ 
35436 & 3.589433 & -30.404231 & 23.10 & 0.3184\tablefootmark{a} \\ 
4136000 & 3.590889 & -30.410893 & 23.11 & 0.2983\tablefootmark{a} \\ 
41419 & 3.577521 & -30.384813 & 23.12 & 0.3219\tablefootmark{a} \\ 
3476000 & 3.585092 & -30.412314 & 23.13 & 0.3071\tablefootmark{a} \\ 
32671 & 3.588878 & -30.413346 & 23.14 & 0.3047\tablefootmark{a} \\ 
35134 & 3.598182 & -30.404825 & 23.15 & 0.3002\tablefootmark{a} \\ 
38748 & 3.579869 & -30.394391 & 23.15 & 0.3201\tablefootmark{a} \\ 
12269000 & 3.577314 & -30.390297 & 23.21 & 0.3131\tablefootmark{a} \\ 
12325000 & 3.605930 & -30.390167 & 23.24 & 0.3059\tablefootmark{a} \\ 
38900 & 3.579626 & -30.394091 & 23.26 & 0.3193\tablefootmark{a} \\ 
42040 & 3.573600 & -30.387182 & 23.27 & 0.2825\tablefootmark{8} \\ 
10930000 & 3.572670 & -30.393633 & 23.28 & 0.3198\tablefootmark{a} \\ 
35978 & 3.596444 & -30.403275 & 23.28 & 0.2989\tablefootmark{a} \\ 
11577000 & 3.590109 & -30.391576 & 23.31 & 0.3080\tablefootmark{a} \\ 
36776 & 3.584729 & -30.399826 & 23.33 & 0.3002\tablefootmark{a} \\ 
3547000 & 3.566721 & -30.411992 & 23.36 & 0.3002\tablefootmark{a} \\ 
33468 & 3.595517 & -30.409682 & 23.42 & 0.3219\tablefootmark{a} \\ 
12191000 & 3.581882 & -30.390474 & 23.44 & 0.3229\tablefootmark{a} \\ 
37304 & 3.583931 & -30.398595 & 23.45 & 0.2986\tablefootmark{a} \\ 
40432 & 3.584083 & -30.388909 & 23.46 & 0.2890\tablefootmark{a} \\ 
34433 & 3.588518 & -30.406316 & 23.47 & 0.3174\tablefootmark{a} \\ 
5079000 & 3.586480 & -30.407004 & 23.54 & 0.3199\tablefootmark{a} \\ 
3795000 & 3.592497 & -30.412103 & 23.55 & 0.3089\tablefootmark{a} \\ 
10265000 & 3.597219 & -30.395534 & 23.56 & 0.3196\tablefootmark{a} \\ 
12018000 & 3.605043 & -30.391671 & 23.64 & 0.2986\tablefootmark{a} \\ 
8216000 & 3.597727 & -30.400602 & 23.69 & 0.3155\tablefootmark{a} \\ 
4938000 & 3.582862 & -30.408838 & 23.70 & 0.3007\tablefootmark{a} \\ 
35340 & 3.588510 & -30.404209 & 23.73 & 0.3198\tablefootmark{a} \\ 
7420000 & 3.568412 & -30.402023 & 23.80 & 0.3054\tablefootmark{a} \\ 
8330000 & 3.569485 & -30.399955 & 23.83 & 0.3049\tablefootmark{a} \\ 
\hline 
14319 & 3.598659 & -30.383098 & 23.38 & -- \\ 
15369 & 3.575794 & -30.380405 & 23.72 & -- \\ 
9490 & 3.574867 & -30.397183 & 23.80 & -- \\ 
10127 & 3.566005 & -30.393924 & 21.12 & -- \\ 
12163 & 3.607988 & -30.390726 & 23.14 & -- \\ 
14786 & 3.593635 & -30.382101 & 23.16 & -- \\ 
5889 & 3.578093 & -30.405740 & 23.76 & -- \\ 
15699 & 3.594368 & -30.378664 & 22.52 & -- \\ 
15508 & 3.578849 & -30.379423 & 23.88 & -- \\ 
7737 & 3.597161 & -30.402246 & 23.97 & -- \\ 
13676 & 3.568877 & -30.385314 & 20.36 & -- \\ 
15624 & 3.575140 & -30.378524 & 21.58 & -- \\ 
5421 & 3.606797 & -30.405592 & 20.39 & -- \\ 
14659 & 3.601322 & -30.382501 & 23.43 & -- \\ 
2937 & 3.580614 & -30.414102 & 22.85 & -- \\ 
15691 & 3.571676 & -30.379704 & 23.55 & -- \\ 
9592 & 3.607186 & -30.397230 & 23.56 & -- \\ 
2818 & 3.576712 & -30.414622 & 23.56 & -- \\ 
16198 & 3.573701 & -30.376807 & 22.99 & -- \\ 
5315 & 3.581605 & -30.407436 & 23.58 & -- \\ 
3313 & 3.575661 & -30.413496 & 23.61 & -- \\ 
15308 & 3.570284 & -30.380798 & 23.64 & -- \\ 
14680 & 3.570831 & -30.382935 & 23.67 & -- \\
\hline
\end{tabular}

\end{table}

\end{appendix}

\end{document}